\documentclass{article}
\usepackage{amsfonts}

\title{\textbf{The Emergence and Evolution of Integrated Worldviews}}
\author{\\Liane Gabora\\\small{Department of Psychology, University of British Columbia}\\\\
Diederik Aerts\\\small{Center Leo Apostel for Interdisciplinary Studies and 
Department of Mathematics}\\\small{Free University of Brussels}}
\date{}
\usepackage[pdftex]{graphicx}
\DeclareGraphicsExtensions{.pdf,.png,.jpg}
\usepackage{lineno}

\begin{document}
\maketitle

\noindent{Reference: 
\\\\
Gabora, L. \& Aerts, D. (2009). A model of the emergence and evolution of integrated worldviews.  {\it Journal of Mathematical Psychology, 53,} 434-451.\\
\\\\

\noindent{Address 
for correspondence:}
\\\\
Liane 
Gabora\\
liane.gabora@ubc.ca\\
Department of Psychology\\
University of British Columbia\\
Okanagan Campus, 3333 
University Way\\
Kelowna BC, V1V 1V7, 
CANADA

\newpage
\noindent{\textbf{\large{ABSTRACT}\\\\}}
It 
is proposed that the ability of humans to flourish in diverse 
environments and evolve complex cultures reflects the following two 
underlying cognitive transitions. The transition from the 
coarse-grained associative memory of \textit{Homo habilis} to the 
fine-grained memory of \textit{Homo erectus} enabled limited 
representational redescription of perceptually similar episodes, 
abstraction, and analytic thought, the last of which is modeled as 
the formation of states and of lattices of properties and contexts 
for concepts. The transition to the modern mind of \textit{Homo 
sapiens} is proposed to have resulted from onset of the capacity to 
spontaneously and temporarily shift to an associative mode of thought 
conducive to interaction amongst seemingly disparate concepts, 
modeled as the forging of conjunctions resulting in states of 
entanglement. The fruits of associative thought became ingredients 
for analytic thought, and \textit{vice versa}. The ratio of 
associative pathways to concepts surpassed a percolation threshold 
resulting in the emergence of a self-modifying, integrated internal 
model of the world, or worldview.

\newpage
\section{Introduction}
\indent 
What enabled humans to flourish in diverse environments, and give 
rise to a process of cultural evolution that is not just cumulative 
but complex, adaptive, and open-ended? Various hypotheses have been 
put forward, such as that it was due to onset of language, tool use, 
or organized hunting. In the paper it is proposed that all of these 
abilities were made possible through two underlying cognitive 
transitions. \\
\indent Like other species, our early ancestors, 
\textit{Homo habilis}, could learn, remember, and perhaps even form 
concepts, but their behavior is much more stereotyped and brittle 
than our own. Donald (1993) refers to their minds as 
\textit{episodic} because they could encode episodes of experience, 
and were sensitive to their significance, but could not voluntarily 
access them. The episodic mind can coordinate appropriate actions in 
response to stimuli, but does not use symbols and makes limited use 
of concepts, and does not enact events, or refine skills. Therefore 
its mental life rarely strays from the `here and now'. 
 \textit{Homo 
erectus} appeared approximately 1.55 MYA and may have co-existed with 
\textit{Homo habilis} for as long as a half million years (Spoor et 
al., 2007), but its brain was substantially larger  (Bickerton, 1990; 
Corballis, 1991; Lieberman, 1991). This period is also associated 
with the origin of culture, as evidenced by such things as the onset 
of tools, organized hunting strategies, and deliberate use of fire. 
Donald refers to the mind of  \textit{Homo erectus} as 
\textit{mimetic} because it could spontaneously recall events without 
a cue, and act them out for others. One thing could now stand for or 
signify another, thus communication took on a semiotic character. It 
could also forge associations, and refine ideas by modifying internal 
representations, a process referred to as \textit{representational 
redescription} (Karmiloff-Smith, 1990, 1992). Its stream of 
experience thereby strayed from the `here and now' into the realm of 
abstractions.\\
\indent A second sudden period of encephalization 
occurred between 500,000 and 200,000 (Aiello, 1996) or 600,000 and 
150,000 years ago (Ruff, Trinkaus, \& Holliday, 1997). Some time later, 
approximately 60,000 years ago, we see the sudden onset of evidence of the 
creativity characteristic of the \textit{modern mind} of \textit{Homo 
sapiens}, such as task-specific tools, art, jewelry, and signs of 
ritualized religion (Bar-Yosef, 1994; Klein, 1989; Leakey, 1984; 
Mellars, 1973, 1989a, b; Mithen, 1996, 1998; Soffer, 1994; Stringer 
\& Gamble, 1993; White, 1993). The modern mind can form abstract 
concepts, combine information from different domains (as in 
analogical reasoning), adapt views and actions to new circumstances, 
and communicate using the complex syntax and recursive embedding 
characteristic of modern human languages. It can frame new 
experiences in terms of previous ones, solve problems using whatever 
potentially relevant information it can obtain, and formulate plans 
of action that reflect the specifics of a situation. In short, a 
modern human behaves as if items in memory are integrated into what 
we will refer to as a \textit{worldview} that provide a big picture 
of what is going on. Its mind is much more than a collection of 
isolated memories, concepts, attitudes, and so forth; it is a manner 
of navigating them, weaving narratives with them, and thereby better 
understanding and interacting with the world. \\
\indent However, the 
existence of this kind of integrated structure presents a paradox. 
Clearly a worldview takes shape in the course of, not just direct 
experience, but streams of thought. For example, by contemplating the 
pros and cons of a democracy one clarifies one's views on democracy. 
However, streams of thought draw upon abstract concepts and their 
associations to mediate how one thought gives way to the next. How 
could one engage in a stream of thought before concepts are woven 
into an integrated structure, and how could concepts be woven into a 
integrated structure prior to the capacity for a stream of thought? 
How could something composed of complex, mutually dependent parts 
come to be? We have a chicken and egg problem. Moreover, what kind of 
thing could concepts be given that their properties and meanings 
shift each time they are called upon? As Bruza \textit{et al.} (this 
volume) note, their behavior defies the conditions of classical 
object-hood and the notion 
 of objectivity.\\
\indent The scenario 
we sketch out for how this could have happened is the following. 
First, we draw upon the notion of autopoiesis as a process that gives 
rise to structure that is self-organizing, self-preserving, and 
self-referential (Maturana \& Varela, 1980; Thompson, 2007). 
Specifically it has been suggested that the chicken and egg problem 
can be resolved in terms of a process of \textit{conceptual 
integration} (Gabora, 1998, 1999, 2000), and in this paper the notion 
is developed more formally. The idea is inspired by the notion that a 
major bottleneck to modern cognition was the evolution of 
self-triggered memory (Donald, 1991). We propose that self-triggered 
thought is made possible by a memory that is sufficiently 
fine-grained that episodes are encoded in rich enough detail to 
regularly evoke associations and remindings. We propose that 
reminding events induce concept formation, which facilitate chains of 
associations, since any concept is associatively linked to more than 
one instance of it, the number of associative paths increases faster 
than the number of concepts. Graph theory tells us that when this 
happens, a percolation  threshold is reached, at which point the 
probability rises sharply that they self-organize into a connected 
closure space (Erdos \& Renyi, 1959, 1960). At this point there 
exists a possible direct or indirect self-triggered associative 
pathway starting from any one memory or concept to any other, and a 
relationally structured worldview thereby emerges. But in order to 
model this we have to tackle the second problem---the extreme 
contextuality of concepts once the worldview starts to become 
integrated. The proposed model grew out of work on the 
State COntext Property or SCOP theory of concepts, which has been used to model how concepts undergo a change of state when acted 
upon by a context, and how they combine (Aerts \& Gabora, 2005a, b; Gabora \& Aerts, 2002; 
Gabora, Rosch \& Aerts, 2008; for work in a similar vein see Bruza \& 
Cole, 2005; Busemeyer et al. 2006; Busemeyer \& Wang, 2007; Nelson \& 
McEvoy, 2007; Widdows, 2003; Widdows \& Peters, 2003). The present paper investigates how SCOP can be used to describe more elaborate conceptual integration such as is essential for a coherent worldview. Although the controversy surrounding 
the issue of modularity is relevant, it is not the focus of this 
paper. The goal here is to show how conceptual integration can be 
described drawing upon insights of the application of the quantum formalisms for the modeling of the conceptual aspects of a worldview, noting relevance 
for the theory that the worldview constitutes the basic unit of 
evolution of culture. 

\section{State COntext Property Theory of 
Concepts}
This section briefly reviews the State COntext Property (SCOP) 
theory of concepts, which provides the mathematical foundation for 
the paper. We emphasize that it is a theory of concepts, not words. Thus when we use SCOP to describe how humans began combining concepts, the ordering of the words used to refer to the concepts (e.g. PET FISH) simply reflects the rules of the English language. \\
\indent SCOP is an elaboration of the State Property formalism 
(Beltrametti \& Cassinelli, 1981; Aerts, 1982, 1983, 1999, 2002). 
SCOP explicitly incorporates the context that evokes a concept and 
the change of state this induces in the concept into the formal 
description of a concept. With SCOP it is possible to describe 
situations with any degree of contextuality. In fact, classical and 
quantum come out as special cases: quantum at the one end of extreme 
contextuality and classical at the other end of extreme lack of 
contextuality (Piron, 1976; Aerts, 1983). The rationale for applying 
it to concepts is expressly that it allows incorporation of context 
into the model of an entity.\footnote[1]{SCOP is a general 
mathematical structure that grew out of (but differs from) the 
quantum formalism, where the context (which is in this case a 
measurement) is explicitly incorporated into the theory. Note that 
this kind of generalization and re-application of a mathematical 
structure has nothing to do with the notion that phenomena at the 
quantum level affect cognitive processes.}

\indent Using the SCOP 
formalism, a concept is described by five elements:
\begin{itemize}
 
	\item A set $\Sigma =$ \textit{\{p, q,  ...\}} of states the 
concept can assume.
	\item A set ${\cal M}$ = \textit{\{e, f, 
...\}} of relevant contexts. (Note that contexts can be concepts.)
 
	\item A set ${\cal L} =$ \textit{\{a, b, ...\}} of relevant 
properties or features. 
	\item A function $\nu$ that describes 
the \textit{weight} (or renormalized applicability) of a certain 
feature given a specific state. For example, $\nu$\textit{(p, a)} is 
the weight of feature \textit{a} for the concept in state \textit{p}. 
Mathematically, $\nu$ is a function from the set $\Sigma \times {\cal 
L}$ to the interval $[0, 1]$. We write 
\begin{eqnarray}
\nu: \Sigma 
\times {\cal L} &\rightarrow& [0, 1] \\
(p, a) &\mapsto& \nu(p, a) 
\nonumber
\end{eqnarray}
	\item A function $\mu$ that describes 
the transition probability from one state to another under the 
influence of a particular context. For example, $\mu$\textit{(q, e, 
p)} is the probability that state \textit{p} under the influence of 
context \textit{e} changes to state \textit{q}. Mathematically, $\mu$ 
is a function from the set $\Sigma \times {\cal M} \times \Sigma $ to 
the interval $[0, 1]$, where $\mu(q, e, p)$ is the probability that 
state $p$ under the influence of context $e$ changes to state $q$. We 
write: 
\begin{eqnarray}
\mu: \Sigma \times {\cal M} \times \Sigma 
&\rightarrow& [0, 1]  \\
(q, e, p) &\mapsto& \mu(q, e, p) 
\nonumber
\end{eqnarray}
\end{itemize}
\noindent Properties in SCOP 
may be either subsymbolic microfeatures or higher-level features that 
have a structure of natural relations as described below in section 
\ref{sec:naturalrelations} (imposed at either the perceptual level or 
the cognitive level). While in some theories (e.g. G\"ardenfors, 2000) context is modeled as a weighting function across attributes or properties, in SCOP any effect of context occurs by way of its effect on the state.\footnote[2]{The inspiration for this choice comes from the physics theories such as quantum mechanics. The primary reason we make this choice is that it is a more ontologically accurate reflection of what a context is and of what a state is. A context has an existence independent of the properties of the concept in question, which is not captured by modeling it as a weighting function across properties.}
A context may consist of a perceived stimulus 
or component(s) of the environment, but it is not necessarily the 
case that the external world figures prominently in the context by 
which one state gives way to the next. The context may consist 
entirely of elements of the associative memory or worldview. In other 
words, the change of state $p \in \Sigma$ to state $q \in \Sigma$ 
under the influence of context $e$---hence for which $\mu(q, e, p)$ 
is the probability---can be internally driven, externally driven, or 
a mix of both. The value of $\mu(q, e, p)$ depends on such factors as 
the extent to which $p$ and $q$ are causally related, share 
properties or conceptual structure at various levels of abstraction, 
as well as spatiotemporal contiguity of relevant episodes in memory, 
their relationship to goals and desires, and so forth. Since episodes 
are understood (and encoded in memory) in terms of the concepts they 
activate, episodes can be said to have properties and undergo changes 
of state due to a context just like concepts. \\
\indent Our minds 
are able to construct a multitude of imaginary, hypothetical, or 
counterfactual deviations from the more prototypical states of 
particular concept, and SCOP can model this. Generally it is not 
possible to incorporate all of the possible contexts that could 
influence the state of a concept. The more states and contexts 
included, the richer the model becomes. The level of refinement is 
determined by the role the model is expected to 
play.\footnote[3]{This is the same methodology as that used to 
describe the (usually infinite) number of different states for a 
physical system. The paradox, namely that it is impossible to 
incorporate all potential contexts into the description, is resolved 
here as it is in physics. For concrete mathematical models, one 
limits the description to a well-defined set of the most relevant 
contexts, hence a corresponding well defined set of states for the 
entity. But in principle it is possible to refine the model 
indefinitely, and hence incorporate ever more contexts and states.} 
What is relevant for our purposes is that unlike other mathematical 
models of concepts, the potential to include this richness is present 
in the formalism, i.e., it can incorporate even improbable states, and 
largely but not completely irrelevant contexts. 

\subsection{Ground States}
An important notion in SCOP is the \textit{ground state} of a 
concept, denoted $\hat{p}$. This is the `raw' or `undisturbed' state 
of a concept; the state it is in when it is not being evoked or 
thought about, not participating in the structuring of a conscious 
experience (such as, most likely, the concept ZEBRA when you began 
reading this paper). 
The ground state is the state of being 
not disturbed at all by the context.
One never experiences a concept in its ground state; it is always 
evoked in some context. The ground state is a theoretical construct; 
it cannot be observed directly but only indirectly through how the 
concept interacts with various contexts (which may include other 
concepts). This is analogous to the fact that a physical system is 
never in empty space. Just as the idealized notion of a physical 
system in empty space is the basis of the description of a real 
physical system in physics, the ground state of a concept plays a 
fundamental role in the SCOP approach to concepts. The properties 
that are actual in the ground state are the characteristic or context-independent (Barsalou, 1982) properties 
of the concept. (The notion of `ground state' is somewhat similar to the notion of `prototype'.)

\subsection{Eigenstates, Potentiality States, and 
Collapse}
\label{sec:eigen}
We saw how, inspired by how the problem of context is approached in physics, the SCOP approach to concepts incorporates the notion `state of a concept', \textit{i.e.} for any 
concept there exists a potentially infinite number of possible states 
it can be in depending on the context that elicits it. Consider a 
concept described by a SCOP $(\Sigma, {\cal M}, {\cal L}, \mu, \nu)$. 
We say that $p \in \Sigma$ is an \textit{eigenstate} of the context 
$e \in {\cal M}$ iff $\mu(p, e, p) = 1$. For example, if state $p$ of 
the concept CLOTHING is an eigenstate of the context $e$, `worn to 
the office' this means that this context $e$ does not cause this 
particular state to change to another state of CLOTHING. The context can be said to be irrelevant. This would be the case if $p$ is the state OFFICE CLOTHING, and thus the context 
`worn to the office' has no effect on it. Note that a state is only 
an eigenstate with respect to a particular context. Under a different 
context the same state may not be an eigenstate.\\

\indent If a state is not an eigenstate with respect to a particular context, then it is 
a potentiality (superposition-like) state for this context, 
reflecting its susceptibility to change. For example, given the context `worn to the office', the  ground state of the concept CLOTHING is a potentiality (superposed) state. Indeed the context `worn to the office' changes this ground state in the state OFFICE CLOTHING.

If $p \in \Sigma$ is a 
\textit{potentiality state} of the context $e \in {\cal M}$ then 
$\mu(p, e, p) < 1$. Same as with eigenstates, a state is only a 
potentiality state with respect to a particular context. A 
potentiality state that is subject to change under the influence of a 
particular context is not necessarily subject to change under the 
influence of any other context. Thus it is through the effect of 
\textit{e} on \textit{p} that sensitivity to context is incorporated, 
and this is what enables a SCOP model of a concept to be the rich, 
imagistic, unbounded mental construct that research has revealed 
concepts to be. Borrowing terminology from quantum mechanics we refer 
to this change of state as \textit{collapse}.\\
\indent Much as 
properties of a quantum entity do not have definite values except 
with respect to a measurement with which they are compatible, 
properties of a concept do not have definite applicabilities except 
with respect to a context with which they are compatible. Let us 
examine how this is dealt with, first in quantum mechanics and then 
in SCOP.\\

\indent 
If a quantum entity prior to a measurement is in an 
eigenstate of this measurement, then the measurement does not change 
this state (it just `detects' what is `in acto'). If the quantum 
entity, prior to the measurement, is in a superposition state with 
respect to this measurement, then the measurement changes this 
state---collapse---and also changes what is actual and what is 
potential. Some properties that were actual become potential and 
\textit{vice versa}. In the formalism of quantum mechanics, a 
superposition state is not an absolute type of state in the sense 
that it is only defined with respect to a measurement. Concretely 
this means that if a certain measurement is performed, then with 
respect to this measurement, each possible state of the quantum 
entity is either a superposition state or an eigenstate. If it is a 
superposition state, this means that the quantity being measured is 
potential (\textit{i.e.} does not have a specific value). If it is an 
eigenstate, then the quantity being measured is actual (does have a 
specific value). The effect of the measurement is to change a 
superposition state to an eigenstate, hence to make quantities that 
were potential before the measurement actual. Quantities that were 
actual with respect to another measurement before the considered 
measurement can become potential with respect to this other 
measurement after the considered measurement. In other words, a 
measurement changes superposition states with respect to this 
measurement into eigenstates, and eigenstates with respect to other 
measurements into superposition states.\\
\indent Similarly, if a 
concept, prior to the context, is in a state of potentiality with 
respect to this context, then the context changes this state 
(collapse), and also changes what is actual as well as what is 
potential. For example, consider the concept TABLE, and two contexts: 
(1) `is laid out with food', and (2) `is cleaned off'. The property 
`is laid out with food' is actual for TABLE in a state that is an 
eigenstate of the first context. In an eigenstate of the second 
context, as the food is cleared away, this property becomes 
potential. Another property, `is cleaned off', which was potential 
for an eigenstate of TABLE in the first context, becomes actual for 
an eigenstate in the second context. It is this dynamics of 
actualities becoming potential and potentialities becoming actual 
that is described using the quantum formalism. The influence of 
context on the state of a concept can be such that even 
characteristic properties of a concept disappear if the concept 
transforms into a new state under the influence of a context. For 
example, consider the concept ISLAND. The property `surrounded by 
water' is a characteristic property, indeed actual in the ground 
state of ISLAND. But if we apply the context \textit{kitchen} to 
island, and hence consider the concept KITCHEN ISLAND, it does not 
have the property `surrounded by water' as an actual property (or 
hopefully not).

\subsection{The Structure of Natural 
Relations}
\label{sec:naturalrelations}
Basic level categories 
(\textit{e.g.} CAT) mirror the correlational structure of properties 
as they are perceived, learned, and used by individuals, and further 
discrimination occurs at the subordinate (\textit{e.g.} TABBY) and 
the superordinate (ANIMAL) levels (Rosch, 1978). The structure of 
this conceptual hierarchy is modeled in SCOP by deriving the 
structure of natural relations for sets of states $\Sigma$, contexts 
${\cal M}$, and properties ${\cal L}$ of a concept (Aerts \& Gabora, 
2005a). Concretely this means the following. If a state \textit{p} is 
more constrained than a state $q$ (\textit{e.g.} TABBY is more 
constrained than CAT) we say that $p$ is `stronger than or equal to' 
$q$, thereby introducing a partial order relation in the set of 
states $\Sigma$. We denote the relation `is stronger than or equal 
to' with the symbol $\le$. We do the same for the set of contexts 
${\cal M}$. Thus if a context $e$ is more constrained than a context 
$f$ (\textit{e.g.} `in the water and near a fallen branch' is more 
constrained than `in the water') we say $e \le f$, thereby 
introducing a partial order relation in ${\cal M}$. By saying this is 
a partial order relation means concretely that it satisfies the 
following mathematical rules. For arbitrary contexts $e, f, g\in 
{\cal M}$ we assume
\begin{eqnarray}
{\rm reflexivity}: \quad && e \le 
e \label{eq:reflexivity} \\
{\rm transitivity}: \quad && e \le f,\ f 
\le g\ \Rightarrow\ e \le g \label{eq:transitivity} \\
{\rm 
symmetry}: \quad && e \le f, \ f \le e\ \Rightarrow\ e = f 
\label{eq:symmetry}
\end{eqnarray}

\indent Next we incorporate more of 
the complexity of the natural world (and our imperfect internal 
mirrors of it) into the model by introducing the `and' context, 
denoted $\wedge$, and the `or' context, denoted $\vee$, corresponding 
to conjunction and disjunction respectively. Thus, given contexts 
$e$, `in the water', and $f$, `near a fallen branch', we can 
construct the context $e \wedge f$ `in the water and near a fallen 
branch'.  Similarly, we can construct the context $e \vee f$, `in the 
water or near a fallen branch'. Elsewhere it is shown that by adding 
the `and' and `or' contexts, ${\cal M}$ obtains the structure of a 
lattice (Aerts \& Gabora, 2005a; see also Aerts, Aerts \& Gabora, 
this issue; Widdows \& Peters, 2003).\\ 
\indent Next we introduce 
the `not' of a context. Given the context $e$, `in the water', the 
context `not in the water'  is denoted $e^\perp$. Technically we say 
that through this introduction of the `not' context, an 
orthocomplementation structure is derived for ${\cal M}$. An 
orthocomplemented lattice structure is also derived for the set of 
properties ${\cal L}$, making it possible to construct a topological 
representation of a SCOP in a closure space. (The closure in the closure 
space appearing naturally at the level of a SCOP of one concept is 
not straightforwardly linked to the notion of closure that we employ 
in the theory of an integrated worldview developed in this paper.)\\

\subsection{SCOP in Hilbert Space}
The dynamics of concepts and how 
they are influenced by contexts can be more precisely modeled by 
embedding SCOP in the complex Hilbert space of quantum mechanics 
(Aerts \& Gabora, 2005b). This enables the weights  of properties to 
be used to calculate the typicality of exemplars or instances. So for 
example, a particular outfit may be seen as more typical of one state 
of  CLOTHING---say, CASUAL CLOTHING---than of another state of 
CLOTHING---say WORK CLOTHING. But with respect to the present paper, 
what is most useful about the Hilbert space formulation is that it 
facilitates the modeling of how states give way or transform into 
other states. Consider two states $p$ and $q$ of a concept. If these 
states are not orthogonal, then following the dynamics of state 
changes implied by Hilbert space, this means that a context $e$ with 
eigenstate $p$, can undergo a change of state from $p$ to $q$. But 
even if this is not possible because states $p$ and $q$ are 
orthogonal it is possible to find another context $f$, such that a 
collapse from $p$ to $q$ occurs by way of this other context. 
Concretely, first context $f$ changes $q$ into one of the eigenstates 
of context $f$, for example $s$. (In Hilbert space $f$ can always be 
chosen such that $s$ is not orthogonal to $p$.) Then, after context 
$f$ has done its work, context $e$ changes $s$ to $p$. The overall 
effect of both contexts $f$ and $e$ is to allow a non-zero 
probability of changing $q$ to $p$, and this is true for any two 
states $p$ and $q$. This shows that Hilbert space describes a 
dynamics of states that are closed, in the sense that one can always 
be changed into another by applying two contexts one after the 
other.\\
\begin{figure} \label{fig:vector}
\begin{center}
 
	\includegraphics[scale=.6]{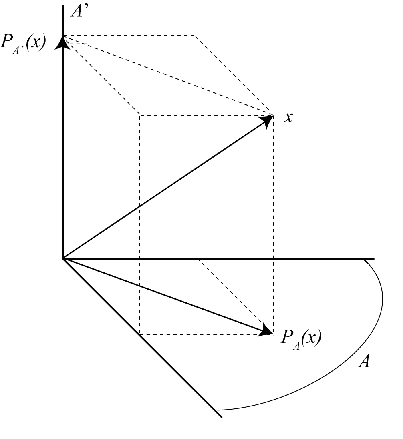}
\end{center}
\caption 
{Graphical depiction of a vector $x$ representing the state of a 
concept, a subspace $A$ representing a context, and its orthogonal 
subspace $A^{\prime}$ representing the negation of this context. 
Under the influence of context $A$, state $x$ collapses to the state 
represented by the orthogonal projection vector $P_A(x)$. Under the 
influence of $A^{\prime}$, $x$ collapses to the state represented by 
the orthogonal projection vector 
$P_{A^{\prime}}(x)$.}
\end{figure}
\indent When a concept interacts 
with a context in Hilbert space, it is immediately projected out of 
the ground state to another state. The change of state induced by a 
context on a concept is graphically represented in Figure 1 by a 
vector $x$, a subspace $A$, its orthogonal subspace $A^{\prime}$, and 
the orthogonal projections $P_A(x)$ and $P_{A^{\prime}}(x)$ on both 
subspaces. The vector $x$ represents the state of a concept, and the 
subspace $A$ represents a context, while the subspace $A^{\prime}$ 
represents the negation of this context. Under the context 
represented by $A$, the state of the concept represented by $x$ 
collapses to the state represented by the projected vector $P_A(x)$, 
and under the negation of this context the state of the concept 
collapses to $P_{A'}(x)$, which is the orthogonally projected vector 
$x$.\\

\subsection{Describing Multiple Concepts with SCOP in Fock Space}
\label{sec:multipleconceptsHibert}

Starting from descriptions 
of separate concepts in a complex Hilbert space SCOP of standard 
quantum mechanics, it is possible to generate a description of 
multiple concepts (Aerts \& Gabora, 2005b, Aerts 2007,  2009). In quantum mechanics the joint entity of multiple entities is described in the Hilbert space that is the tensor product of the Hilbert spaces describing these entities. However, it turns out that the quantum field theory procedure to describe multiple concepts gives rise to a more accurate model of what happens when concepts combine than the tensor product. Fock space is the direct sum of tensor products of Hilbert spaces, so it is also a Hilbert space. It is the fact that it is the direct sum of tensor products of Hilbert spaces that gives Fock space the kind of internal structure that makes it useful for the modeling of multiple concepts and their combinations (Aerts, 2007, 2009). To summarize briefly, suppose we consider a collection of $n$ concepts 
\begin{equation}
{\cal U}=\{u_1, \ldots, u_j, \ldots, u_n\}
\end{equation}
and also a combination of these $n$ concepts. 
First we introduce the complex Hilbert space ${\cal H}$ that describes all of these concepts, such that the state of each concept $u_j$ is represented in this Hilbert space by mean of a unit vector $|u_j\rangle \in {\cal H}$.

Before we model the general situation, let us model the situation that the combination constitutes a new concept. The state $|u\rangle_{new}$ of this new concept is a weighted linear combination of the states of all the concepts.
\begin{equation}
|u\rangle_{new}=\sum_{j=1}^na_je^{i\alpha_j}|u_j\rangle
\end{equation}
with $0 \le a_j \le 1$ and $\sum_{j=1}^na_j^2=1$ and $e^{i\alpha_j}$ the phase of each of the complex numbers $a_je^{i\alpha_j}$, where we have written the complex numbers of the linear combination in polar form.
There is another special situation, namely that the combination is a collection of $n$ distinct and separate concepts, hence the opposite of coming together as `one' new concept. The state $|u\rangle_{col}$ of such a collection of concepts is the tensor product of the states of all concepts.
\begin{equation}
|u\rangle_{col}=\otimes_{j=1}^n|u_j\rangle
\end{equation}
All the intermediate situations are also possible, of course, e.g. where part of the collection of concepts is treated as a new concept while others remain distinct. Let us mathematically express this general situation, hence also including the intermediate ones. First we write the state of the situation where for $0 \le k \le n$, there are $k$ subsets $\{{\cal U}_j\ \vert j=1\ldots k\}$ of ${\cal U}$, hence ${\cal U}_j \subseteq {\cal U}\ \forall j$ and ${\cal U}=\cup_{j=1}^k{\cal U}_j$ and ${\cal U}_{j_1} \cap {\cal U}_{j_2}=\emptyset$ for $j_1\not=j_2$, such that for each of these subsets there is a new concept. The state of this concept is then the weighted linear combination of the states contained in that subset. Since we have considered $k$ such subsets, the global state of all of the collection of concepts is then a vector of $\otimes_{j=1}^k{\cal H}_j$, the $k$-times tensor product of the Hilbert space ${\cal H}$. This means that the state of the collection is then the weighted superposition of each of these states, which is an element of the direct
%% start change L
sum
%% end change L
for all values of $k$ of  $\otimes_{j=1}^k{\cal H}_j$, hence of the space $\oplus_{k=1}^n\otimes_{j=1}^k{\cal H}_j$, which is Fock space. A general state $|u\rangle_{gen}$ of this Fock space can be written as
\begin{equation} \label{generalstate}
|u\rangle_{gen}={1 \over N}\sum_{k=1}^nm_ke^{i\theta_k}(\sum_l a_{lk}e^{i\alpha_{lk}}(\otimes_{j=1}^k|v_{jl}\rangle))
\end{equation}
where $N$ is a normalization constant such that $\langle u_{gen}|u_{gen}\rangle=1$, and $|v_{jl}\rangle \in {\cal H}$ are states of the individual concepts, however not necessarily of the forms $|u_j\rangle$.
%% start change LL
Some of them might be of this form; others might be weighted superpositions of such states, if they express combinations of parts as a whole, and others might be orthogonal projections of such states, if the effects of contexts are taken in account explicitly in a combination. The sum $\sum_l a_{lk}e^{i\alpha_{lk}}(\otimes_{j=1}^k|v_{jl}\rangle)$ expresses that we consider also entangled states of such $k$ tensor product states $\otimes_{j=1}^k|v_{jl}\rangle$ as possible states of that part of the general state of the combination of concepts described in sector $k$ of Fock space.
%% end change LL

Consider an example, namely the conception of a child playing in the garden. The individual concepts are CHILD, PLAY, and GARDEN. Each of them is described by a ket-vector, hence we have the set of vectors ${\cal U}=\{|u_1\rangle, |u_2\rangle, |u_3\rangle\}$, where $|u_1\rangle$ is the state of CHILD, $|u_2\rangle$ is the state of PLAY, $|u_3\rangle$ is the state of GARDEN. 
The state denoted $|u\rangle_{new}$ is the weighted superposition of the three states that represent the combination as one new concept. The state denoted $|u\rangle_{col}$, is the $3$ tensor product state of the 3 states. It represents them as a collection of separate concepts. The intermediate situations are those for which subsets of the collection of concepts are considered new concepts, while these subsets themselves are considered to constitute distinct elements of the whole set. All these situations are described by the state denoted $|u\rangle_{gen}$ of Fock space.

Many subtle aspects of how concepts come together can be modeled by the Fock space representation. 
If two concepts come together as a new conceptual element, then in the Fock space state representation, the linear combination coefficient giving rise to the weight for this factor becomes dominant compared to those for other concepts that do not come together to be considered as something new. There is a dynamical shifting of the weights in Fock space. Thus the quantum modeling scheme contains the fine structure to describe the variety of ways one could react to a particular collection of concepts. This is why we model a cognitive state as the state of Fock space that describes all concepts and combination of concepts of the worldview at a particular instant.

%% start change L
One additional remark is at place here. The Fock space procedure makes it possible to construct mathematically the state of a combination of many concepts. However, some combinations are more easily modeled using a procedure described detail elsewhere (Aerts \& Gabora 2005a,b). These are the combinations where certain concepts are brought into the conceptual structure as contexts rather than as entities. We have incorporated concepts that act as a context in a combination in the Fock space representation, since in equation (\ref{generalstate}) we suppose the considered collection ${\cal U}$ to consists of concepts that are entities in the considered combination, hence not concepts that are contexts. However, we allow concepts that are context to change the states of the entity concepts as described above, and this possibility is contained in equation (\ref{generalstate}). The description put forward here is also fine enough to take into account that the distinction between a concept playing the role of context or rather playing the role of entity within a specific combination can be ambiguous, and can also change depending on the overall context. A good example is STONE LION. Different possible ways of looking at this combination exist, and a positive answer to both questions `is STONE LION a LION' and `is STONE LION a STONE' make sense. Hence a better approach is to consider both as entities, and use Fock space for a two entity situation, and forgo the more simple model of considering one of them as a context. 
%% stop change L

\section{SCOP Applied to Cognitive States}
Until now SCOP has been used to describe one or more 
concepts. This section presents material needed to apply it to the 
entire set of concepts possessed by an individual, as is necessary to 
model the evolution of an integrated worldview. We first discuss the 
structure of associative memory, and the distinction between states 
of a concept and cognitive states, and outline the various kinds of 
internal pathways involved in the transformation or collapse from 
cognitive state to another.

\subsection{Ways of Incorporating 
Associative Structure in Models of 
Cognition}
\label{sec:assocstructure}
It is well-known that memory 
neurons respond to specific properties, or subsymbolic microfeatures 
(Churchland \& Sejnowski, 1992; Smolensky, 1988). For example, one 
might respond to a particular shade of blue, or the quality of 
`stickiness', or quite likely, something that does not exactly match 
an established term (Mikkulainen, 1997). Items are 
\textit{distributed} across a cell assembly that contains many 
neurons, and likewise, each neuron participates in the storage of 
many items. Thus, the same neurons get used and re-used in different 
capacities, a phenomenon referred to as neural re-entrance (Edelman, 
1993). Items stored in overlapping regions are correlated, or share 
features. Memory is also \textit{content addressable}, \textit{i.e.} 
there is a relationship between the semantic content or meaning of an 
item, and which neurons are activated by it and participating in 
encoding it. As a result, episodes stored in memory can thereafter be 
evoked by stimuli that are similar or resonante in some (perhaps 
context-specific) way (Hebb, 1949; Marr, 1969). It follows that the 
likelihood that one item evokes another varies with the extent to 
which they share properties. Most models of concepts include only a 
small number of properties as possible properties for a concept to 
have, which inevitably is too limited to account for what happens 
when a concept appears in a context quite different from previous 
ones, or when it combines with other concepts. As mentioned 
previously, SCOP allows that for each concept there exists some 
context that could actualize a certain property that was thought to 
be very unlikely for that concept; thus each of the properties can 
potentially play a possible role in any concept (though the weights 
for most properties for any given state will tend to equal zero). 
\\
\indent Kanerva (1988) makes some astute observations about the 
conceptual space that can be realized by a sparse, distributed, 
content-addressable memory such as human memory. We denote the number 
of subsymbolic microfeatures, or properties, that are activated by 
some neuron and can be encoded in the memory as $n$. The set of all 
possible $n$-dimensional items that can be encoded in the memory are 
represented as the set of vertices (if features assume only binary 
values) or points (if features assume continuous values) in an 
$n$-dimensional hypercube, where the $s$ items that actually are 
encoded in the memory occupy some subset of these points. The 
distance $d$ between two points in this space is a measure of how 
dissimilar they are, referred to as the Hamming distance. The number 
of items encoded in memory at Hamming distance $d$ away from any 
given item is equal to the binomial coefficient of \textit{n} and 
\textit{d}, which is well approximated by a Gaussian distribution 
(Figure 2). Thus, if item X is 111...1 and its antipode is 000...0, 
and we consider X and its antipode to be the `poles' of the 
hypersphere, then approximately 68\% of the other items lie within 
one standard deviation ($\sqrt{\textit{n}}$) of the `equator' region 
between these two extremes. As we move through Hamming space away 
from the equator toward either X or its antipode, the probability of 
an item falls off sharply by the proportion 
$\sqrt{\textit{n}}/\textit{n}$. \\
\begin{center}
 
	\includegraphics[scale=.9]{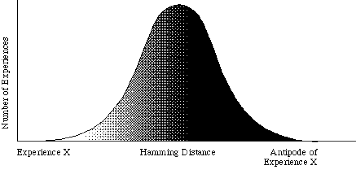}
\end{center}
\noindent 
Figure 2. Solid black curve is a schematic distribution of the 
Hamming distances from any given item to other items encoded in 
memory.  The Gaussian distribution arises because there are many more 
ways of sharing an intermediate number of features than there are of 
being extremely similar or different. Activation decreases with 
distance from the most similar address, as indicated by green 
shading. (A typical computer memory stores each item in only the 
left-most address, whereas a distributed network stores it throughout 
the network. A neural network that uses an activation function, such 
as the radial basis function, is intermediate between these two 
extremes.)\\\\ 
\indent Of course conceptual distance is context sensitive, but not in a chaotic way; thus for
example, ROSEBUD is closer to PANSY than to SPACESHIP across the vast majority of contexts. So the Hamming distances here are for items in their most prototypical forms, i.e., their most typical or commonly encountered contexts.)

Note that Figure 2 depicts not the 
distribution of actual items in memory, but of possible items; with 
respect to actual items encoded in memory, the tail regions would be 
considerably sparser. This means that while there are many ways of 
being somewhat similar to any given item, very few items are either 
very similar to, or very dissimilar to, a given item, and the lower 
$n$ is the more pronounced the effect. This has implications for the 
model developed here because interaction and integration is more 
probable amongst items that are similar (or more correctly, more 
correlated). Langton's (1992) finding that the information-carrying 
capacity of a system is maximized when its interconnectedness falls 
within a narrow regime between order and chaos suggests that to 
produce a steady stream of meaningfully related yet potentially 
creative remindings, the correlation between one thought and the next 
must fall within a narrow range. \\
\indent It is interesting to look 
at how characteristics of associative memory are realized by SCOP as 
compared to a neural network. A neural network is context-sensitive 
in the sense that the output depends on (1) not just one feature but 
all features of an input, (2) what patterns the neural network has 
learned previously (reflected in the weights on links between nodes), 
and (3) the activation function that gives the relationship between 
activation of a node and its output to other nodes. The fact that 
items are distributed in memory is modeled in a neural network 
through parallel distributed activation of nodes. The fact that 
memory exhibits coarse-coding, such that a given neuron responds 
maximally to one microfeature but responds to a lesser degree to 
other similar microfeatures, can be modeled by constraining the 
distribution using a radial basis function (RBF). Each input 
activates a hypersphere of hidden nodes, with activation tapering off 
in all directions according to a (usually) Gaussian distribution of 
with standard deviation $\sigma$ (Willshaw \& Dayan, 1990; Hancock 
{\it et al.}, 1991; Holden \& Niranjan, 1997; Lu {\it et al.,} 1997). 
This enables one part of the network to be modified without 
interfering with the capacity of other parts to store other patterns. 
Where $F$ is the activation, $x$ is an i-dimensional input vector, 
and $k$ is the center of the RBF, hidden nodes are activated as 
follows
\begin{equation}
F(x)=e^{-\sum_i((x_i - k_i / 
\sigma)^{2})}
\end{equation}
The further a node is from $k$, the less 
activation it receives from the input and in turn contributes to the 
output. If $\sigma$ is small, the input activates few neurons but 
these few are hit hard; we say the activation function is {\it 
spikey}. If $\sigma$ is large, the input activates many neurons to an 
almost equal degree; we say the activation function is relatively 
{\it flat}. In effect $\sigma$ moderates the degree to which the 
output is affected by other patterns that have previously been 
encoded.\\
\indent This approach nicely resolves the problem of 
interference amongst stored patterns, but like all neural net 
approaches, it assumes the validity of the notion of spreading 
activation, which has recently been seriously called into question 
(Bruza, Kitto, Nelson \& McEvoy, this volume; Nelson, McEvoy \& 
Pointer, 2003; Nelson \& McEvoy, 2007). In SCOP, a similar effect can 
be obtained using a variable we will refer to as the 
\textit{association threshold}, denoted $\phi$, which adjusts the 
extent to which associations constitute part of the context that 
evokes the change of state from $p$ to $q$. For example, at a given 
moment you may be pruning a tree. Let us refer to this cognitive 
state as ${\bf p}$. In your next cognitive state, let's call it ${\bf 
q}$, you might be reminded of another tree you have pruned. This 
change of cognitive state is possible because both ${\bf p}$ and 
${\bf q}$ are instances of the concept TREE-PRUNING with a non-zero 
transition probability $\mu$\textit{(q, e, p)}, The change of 
cognitive state is the result of associative recall (as discussed in 
section 3.2.2), i.e., there exists a pathway between ${\bf p}$ and 
${\bf q}$ denoted ${\bf p} \rightarrow {\bf q}$.\\
\indent 
Alternatively your associations may be not at the level of properties 
but at the conceptual level; you might be reminded of your family 
tree, or the tree of knowledge. You might be hungry or your hand 
might be getting sore. All of these are possible components of the 
context that evokes the collapse from state $p$ of the concept to 
state $q$. The function $\phi$ modulates how you draw upon these various elements 
of your situation to create a specific context by which your current 
thought transforms into your next thought, as follows. If $\phi$ is 
low, how $p$ transforms into $q$ is affected by other episodes 
previously encoded in the neurons activated by $p$. The context that 
evokes the collapse is composed of associations at the property level 
and/or at other higher levels of abstraction. In other words, we 
posit that the direction of one's thought is shaped by a choice of 
contexts that takes associations into account. Since it is impossible 
to model with complete precision how these associations affect the 
collapse from $p$ to $q$ the model contains a lack of knowledge and 
$\mu$ enables us to incorporate this lack of knowledge into the 
model. Lowering $\phi$ allows increasingly distant associations to be 
made; referring to figure 2 it will tend to increase how great  the 
Hamming distance can be (how far into the grey area) between the 
current experience (experience X) and an item in memory in order for 
that item to be evoked by the current experience. If on the other 
hand $\phi$ is high, associations play little role; in other words, 
how $p$ transforms into $q$ is not affected by what other episodes 
have been encoded in the neurons activated by $p$. Since mental 
energy is not going toward finding associations it is freed up for 
symbolic operations on $p$. Since associations do not affect the 
collapse from $p$ to $q$ there is much less lack of knowledge, hence 
$\mu$ values tend to be 0 or 1.

\subsection{Cognitive States versus States of a Concept}
A cognitive state of the mind of an individual 
is a state of the composition of all of concepts and combinations of 
concepts of the worldview of this individual.
Referring to the mathematical representation for a large combination of concepts that we put forward in section \ref{sec:multipleconceptsHibert}, we can consider the state representing the combination of all concepts contained in the worldview of an individual, i.e., the cognitive state of the mind of this individual. Hence, making use of equation (\ref{generalstate}), and denoting by ${\cal U}$ the collection of concepts available to an individual, where we denote the state of an concept $u \in {\cal U}$ by means of the ket-vector $|u\rangle \in {\cal H}$, where ${\cal H}$ is a complex Hilbert space, this gives
\begin{equation}
{\bf p}={1 \over N}\sum_{k=1}^nm_ke^{i\theta_k}(\sum_l a_{lk}e^{i\alpha_{lk}}(\otimes_{j=1}^k|v_{jl}\rangle))
\end{equation}
where $N$ is a normalization constant such that $\langle {\bf p}|{\bf p}\rangle=1$, and $|v_{jl}\rangle \in {\cal H}$ represent states of the individual concepts, however not necessarily of the forms $|u\rangle$ for an $u\in{\cal U}$. Some of them might be of this form, others might be weighted superpositions of such vectors,
%% begin change L
and still others might be normalized projections of such vector due to concepts that only function as context in the considered combination (we have not included such concepts in ${\cal U}$).
%% end change L
For a given 
cognitive state ${\bf p}$, most concepts ${u \in {\cal U}}$ are in their ground 
state. They are not evoked in this instance of experience, hence we 
say they are not activated. Those concepts ${u \in {\cal U}}$ that 
are not in their ground states for the cognitive state can be said to 
be activated, or in the language of quantum mechanics we say they are 
in `excited states'.\\
\indent 

\indent  Note that ${\cal U}$ refers to the totality of what has 
been conceived (framed in conceptual terms), not what has been 
perceived. It is possible that one could have an experience that 
activates memory but such that it is not framed in conceptual terms. 
For example, one might have an experience (such as ``I like the white 
pebble in the water") without verbalizing it as such even to oneself, 
or in the case of an infant without even possessing the concepts 
WHITE, PEBBLE, and WATER. That is, one might simply look at the thing 
and like it. We reserve the term \textit{cognitive states} to refer 
specifically to episodes that activate concepts. A cognitive state 
that is not conceptualized, \textit{i.e.} not interpreted in terms of 
concepts, may still give way to another state through cognitive 
events such as reminding, induction, and so forth, but this change of 
state reflects the set of properties \textit{\{a, b, ...\}} of the 
current episode, and the properties of similar previously encoded 
episodes, not the conceptual structure by which these sets of 
properties have been interpreted and organized. \\
 \indent  We use 
the term {\it episode} to refer to a cognitive state that is either 
being encoded or has been encoded in long-term memory and that can 
thus potentially be recalled from memory (either completely or 
partially, accurately or inaccurately). The content of an episode can 
range from rich perceptual data to a highly abstract or imaginary 
mental construct that has little relation to something ever directly 
perceived. The collection of all episodes is ${\cal X}$, and episodes 
will be denoted $x, y, z \in {\cal X}$.\\
\indent  Change of 
cognitive state is recursive. The new cognitive state is a function 
of the effect on the worldview of (1) the previous cognitive state 
and (2) the perceived world.
% start change L
Because we use the quantum formalism, or its generalizations, to mathematically describe states, contexts, and properties, it follows from this quantum formalism how dynamics should be modeled. One models dynamics by building the Schr\"odinger equation, or the generalized version of it. The Schr\"odinger dynamics models the change that is provoked by the influence of all the contexts. Mathematically we write
\begin{equation}
i{\partial \over \partial t} {\bf p}(t)=H{\bf p}(t)
\end{equation}
\noindent where $H$ is the Hamiltonian, which is a self-adjoint operator in the Hilbert space ${\cal H}$. Let us write the exponential form of the Schr\"odinger equation, which is equivalent with the standard form, but more easy to interpret in the situation where we use it. We have
\begin{equation}
{\bf p}(t)=e^{iH(t-t_0)}{\bf p}(t_0)
\end{equation}
This means that the cognitive state at an arbitrary time $t$ is given by the cognitive state of time $t_0$, transformed under influence of the unitary transformation which is the exponential of $i$ times the Hamiltonian operator
%% start change L
multiplied by the time interval $t-t_0$.
%% stop change L

When quantum mechanics is applied in physics to describe the the micro-world, the Hamiltonian $H$ of a quantum entity under study is the operator corresponding to the energy of the entity. Hence, it is the energy of the quantum entity that determines how its state changes under the joint influence of all contexts. There is no obvious way to define a notion that plays the role of `energy' in our situation where the entity is a worldview instead of a quantum entity. One possibility is to approach this situation the other way around. One looks for a self-adjoint operator $H$, such that the dynamics of a worldview is described well by the Schr\"odinger equation, and then the quantity corresponding to this self-adjoint operator is introduced as being the energy of the worldview.\footnote[4]{The Schr\"odinger equation has been used before in this way to model the dynamics of a cognitive entity. Aerts, Broekaert and Smets (1999a,b) use Schr\"odinger equations to model the cognitive dynamics of the Liar paradox. Busemeyer, Matthew and Wang (2006), use the Schr\"odinger equation in a quantum game theoretic setting to model a decision of the type where the disjunction effect appears, and a Hamiltonian is calculated with respect to experimental data measured in this situation.}
% L

\subsection{Ways of Forging Conceptual Structure}
This section identifies processes through which 
conceptual structure is formed. This includes the formation of 
concepts, and of pathways connecting episodes to episodes, episodes 
to concepts, and concepts to other concepts. In other words, pathways 
provide internally generated means of going from thinking about or 
experiencing one thing to thinking about or experiencing another. The 
strengths of these pathways is reflected in the transition 
probabilities of SCOP. 

\subsubsection{Pavlovian Conditioning and 
Spatiotemporal Contiguity}
One source of pathways arises through 
Pavlovian or classical conditioning, as in the pairing of a bell with 
food. The episode (bell with food) can be said to instantiate 
multiple concepts (BELL and FOOD). At the heart of classical 
conditioning is spatiotemporal contiguity--the pairing of a 
conditioned stimulus with an unconditioned stimulus causes a 
response--an association--to the unconditioned stimulus. However 
neutral stimuli that go together in space and /or time can similarly 
become associated. To say that two or more properties are properties 
of a particular concept or cognitive state is to say that they 
possess associative strucure due to spatiotemporal contiguity, thus 
this kind of structure is built into SCOP. For example, whether or not 
BELL has been paired to FOOD, most instances of BELL involve the 
sight and sound of a bell--two dissimilar properties. As another example the episode `Mary is wearing 
a ball gown' is an instance of the concept MARY denoted $u \in {\cal 
U}$ and an instance of the concept BALL GOWN denoted $v \in {\cal 
U}$. Following the experience of this episode, thoughts of  Mary 
could remember one of ball gowns and thoughts of ball gowns could 
remind one of Mary. We say that there exists a bidirectional pathway 
between $u$ and $v$ denoted $u \leftrightarrow 
v$.

\subsubsection{Associative Recall}
Another source of pathways 
arises through the experience of an episode that is so similar to 
some previously encoded episode that it evokes a reminding or recall 
of that previous episode. For example, the episode `Mary is wearing a 
ball gown' may evoke a reminding of the episode `Jane is wearing a 
ball gown'. Let ${\bf p}$ be the cognitive state of experiencing the 
episode `Mary is wearing a ball gown'  and ${\bf q}$ be the cognitive 
state of experiencing the episode `Jane is wearing a ball gown'. If 
${\bf p}$ evokes ${\bf q}$ we say that there exists a pathway between 
${\bf p}$ and ${\bf q}$ denoted ${\bf p} \rightarrow {\bf 
q}$.

\subsubsection{Abstraction} 
An important source of conceptual 
structure is recognition that particular episodes could be seen as 
states of a new concept, resulting in the formation of that new 
concept through the process of abstraction. This can occur when due 
to the distributed nature of memory, the associative recall process 
retrieves an episode with substantial overlap of features or 
conceptual structure, An example is CAT reminds one of DOG resulting 
in formation of the concept PET. Abstraction generally involves the 
formation of a more abstract concept out of less abstract (more 
constrained) concepts. With SCOP this is described as follows. 
Suppose that $p$ is an episode involving APPLE (and perhaps 
another episode $q$ involving ORANGE). Abstraction occurs if one 
realizes (or is told) that there could exist a concept $u \in {\cal 
U}$, the concept FRUIT represented by the SCOP $(\Sigma_u, {\cal 
M}_u, {\cal L}_u, \mu_u, \nu_u)$  such that  $p \in \Sigma_u$, and 
similarly $q \in \Sigma_u$. This
  can arise spontaneously as a 
result of associative recall; for example an apple reminds one of an 
orange which leads to the formation of an item in memory that 
encompasses both.\\

\subsubsection{Identifying Natural 
Relations}
Pathways are forged through the learning or recognition of 
natural relations as discussed in section \ref{sec:naturalrelations}. 
For example consider the concepts $p$ and $q$. If one learns or 
realizes that $e \wedge f \le e$, a pathway $e \wedge f 
\leftrightarrow e$ is  generated. Similarly if one learns or realizes 
that $e \le e \vee f$, a pathway $e \leftrightarrow e \vee f$ is 
generated.\\
\indent An important kind of pathway is the 
identification of episodes as states or instances of a particular 
concept, or of concepts as states of more abstract concepts. This is 
referred to as {\it instantiation}. With SCOP this is described as 
follows. Suppose that $u \in {\cal U}$ is the concept APPLE, 
represented by the SCOP $(\Sigma_u, {\cal M}_u, {\cal L}_u, \mu_u, 
\nu_u)$, and $p$ is a state of apple, say, FUJI APPLE. If one learns 
or recognizes that $p \in \Sigma_u$, then we say that there is an 
`isa' pathway connecting $p$ with $u$, denoted $p \rightarrow u$, as 
well as a pathway $u \rightarrow p$. \\
\indent If there exists a 
concept $w \in {\cal U}$ such that $p$ and $q$ are both known to be 
states or instances of $w$, hence $p, q \in \Sigma_w$. Thus $p 
\leftrightarrow w$ and $q \leftrightarrow w$. Since $p \rightarrow w$ 
and $w \rightarrow q$, through transitivity, $p \rightarrow q$. In 
such cases we refer to $p$ and $q$ as \textit{sister instances} of 
$w$.\\

\subsubsection{Conjunctions}
It is not always the case that a 
new concept is more abstract or less constrained than the concepts 
that gave rise to it. Another important source of conceptual 
structure is the formation of a conjunction of concepts, as in the 
joining of the concepts SNOW and MAN to give the idea of a man built 
out of snow, a SNOWMAN. This occurs when due to the distributed 
nature of memory, what the associative recall process retrieves has 
components that lack substantial overlap of features or properties, 
and that under {\it most} contexts would not be retrieved together, 
and were thought to be unrelated. In SCOP this is mathematically 
described as in Section \ref{sec:multipleconceptsHibert}. Two 
concepts combine to give a conjunction using the tensor product 
formulation which results in (amongst other states) states of 
entanglement. Consider a joint concept $u_1 \otimes u_2$ described in 
the complex Hilbert space SCOP formed by means of the tensorproduct 
${\cal H}_1 \otimes {\cal H}_2$ of the two Hilbert spaces ${\cal 
H}_1$ and ${\cal H}_2$. If one or more of the states $p \in u_1 
\otimes u_2$ is an entangled state, we say that there exists a 
bidirectional pathway connecting the two concepts that comprise the 
conjunction, hence $u_1\leftrightarrow u_2$. 
% LG This next bit is from what you added much earlier as an 'easier' way of getting new combined concepts without entanglement. Is it still worth keeping it in the paper? 
% LG repeats above question.
% LG Please answer the 
above question. 
% LG asks a 4th time..... same question above. 

%%%%% LG Why do you keep not answering this question?
Alternatively, 
we say there exists a pathway if the following condition is met. 
There exists a concept $v$, such that $v$ in state $q \in \Sigma_v$ 
is a context for concept $u$ in state $p \in \Sigma_u$. We call it 
$e_q \in {\cal M}_u$. $e_q$ changes $p$ to a new state of  concept 
$u$. This state is $q \wedge p \in \Sigma_u$, and $\mu(p \wedge q, 
e_q, p) \approx 1$.\\

\subsection{The Chaining and Integration of Pathways}
% LG3 Can you please formulate the following in mathematical terms?
The above-mentioned ways of forging conceptual 
structure work by way of changing the set ${\cal M}$ of contexts 
associated with a given concept. For example, consider a SCOP that 
consists of the concepts STONE and ANIMAL and instances of ANIMAL, 
such as LION, TIGER, and MONKEY. We model an episode involving 
exposure to a lion made of stone as the concept combination STONE 
LION. The set of contexts associated with other animal concepts 
expands to include the potentiality that the animal can be made of 
stone. Thus ${\cal M}$ for MONKEY now includes STONE MONKEY and 
${\cal M}$ for TIGER now includes STONE TIGER. \\
\indent We said 
that with respect to a concept, the change of state $p \in \Sigma$ to 
state $q \in \Sigma$ under the influence of context $e$---hence for 
which $\mu(q, e, p)$ is the probability---can can be internally 
driven, externally driven, or both. This is also the case with 
respect to the context that drives a change of cognitive state; it 
may consist of a perceived stimulus or component(s) of the 
environment, or elements of the associative memory or worldview, or a 
mixture of the two. Pathways allow the autonomous (hence internally 
driven, not dependent upon change in the external environment) 
transformation of thoughts and ideas. \\
\indent Instances of 
self-triggered recall can be chained together recursively resulting 
in a stream of thought. A change from one cognitive state $u \in 
{\cal U}$ to another $v \in {\cal U}$ may in turn modify the 
worldview, hence the (internally driven component of the) context by 
which $v$ is 'measured' or made sense of. Although $u$ may have been 
an eigenstate with respect to the context provided by the previous 
state of the worldview, $v$ may well not be an eigenstate with 
respect to the context provided by this newly updated version of the 
worldview. This means that there is another collapse or change of 
state from $v$ to $w$. This can continue recursively, until the 
concept enters a state that is an eigenstate with respect to the 
context provided by the worldview due to the effect on it of the 
previous cognitive state. It is through this chaining of episodes 
that indirect pathways or what we will refer to as {\it routes} 
through associative memory are forged. \\
\indent $x, y, z \in {\cal 
X}$ can be said to be {\it integrated} if for each episode there 
exists \textit{some} possible direct or indirect route to any other 
episode by way of pathways. If this is the case the whole exhibits 
emergent structure that cannot be reduced to the sum of the parts.\\
\indent Thus conceptual structure arises by way of the creation of 
new concepts through abstraction and concept combination, and the 
creation of new pathways amongst episodes and concepts through 
processes such as associative recall, spatiotemporal contiguity, and 
the identification of natural relations. Examples of the formation of 
conceptual structure will be elaborated in the tentative model of the 
transition to cognitive modernity proposed in the pages to come.

\section{Modeling the Episodic Mind}
Now that the basic elements of 
SCOP and the extension of it to describe conceptual integration have 
been introduced, let us apply it to the subject of this paper: 
evolution of the capacity to develop an internal model of the world, 
or worldview, the components of which are accessed spontaneously, 
recursively, and in a context-sensitive manner to interpret (and 
generate) elements of culture. We begin by modeling the mind for 
which this is not the case. \\ 

\subsection{Coarseness of Memory 
Limits Associative Recall and Conceptual Structure}
The episodic mind 
may be innately predisposed to categorize some situations---such as 
for example situations involving a member of another species---as 
instances of FRIEND, FOE, OR FOOD (without possessing words for these 
concepts). This facilitates the formation of pathways due to 
spatiotemporal contiguity. An experience that might not have evoked 
an episode encoded in memory if it had {\it not} been encoded as an 
instance of FRIEND might evoke it if it {\it is} encoded as an 
instance of FRIEND because the fact that both are instances of FRIEND 
makes the similarity more salient. Thus innate concepts facilitate 
the formation of pathways through associative recall, and thereby 
allow for a limited development of conceptual structure.  \\
\indent 
Although the episodic mind is capable of associative recall---for 
which the context that causes the change of cognitive state is 
something in the externally perceived world---it cannot engage in 
self-triggered recall---for which the context is generated 
internally. This means that instances of associative recall are not 
chained together recursively to give a stream of self-triggered 
thought. For this reason, the episodic minds of early hominids tended 
not to deviate far from the `here and now' of perceived episodes, as 
opposed to the world of mental constructs and imagination. Why might 
our early ancestors have been incapable of self-triggered recall? It 
is reasonable to assume that the smaller brain of the early hominid 
had fewer neurons, and thus \textit{n} was small; that is, with few 
properties or subsymbolic microfeatures distinguishing one episode or 
item encoded in memory from another. Thus episodes were encoded in 
less detail. Since associative recall requires similarity of features 
or conceptual structure, and since there are fewer features, there 
are fewer routes for associative recall (in which one episode evokes 
another). On those rare occasions when associative recall does occur, 
it is unlikely to occur immediately afterward, and even if it does, 
since the context that causes it is external as opposed to internal, 
the result is not refinement or honing of the current thought. As 
mentioned earlier in the discussion of Figure 2, as we move through 
Hamming space away from the equator toward either $X$ or its 
antipode, the probability of encountering a memory item falls off 
sharply by a maximum (depending on the number and nature of items 
encoded) of $\sqrt{n}/n$. For the episodic mind, the region of 
substantial overlap between $X$ and another episode is extremely 
small.\\
\indent 
The situation is compounded by the fact that since 
two processes by which new concepts come about, abstraction and 
conjunction, depend on associative recall, episodes are less apt to 
be conceptualized, or framed in terms of their conceptual structure. 
Since unconceptualized episodes lack conceptual structure they are 
not as deeply rooted into the worldview; thus there are few possible 
retrieval routes. As a simple example, an episode involving a ball 
but not conceptualized as an instance of TOY would not evoke memories 
of other toys such as dolls because the similarity between balls and 
dolls resided primarily at the conceptual level, not the perceptual 
level. Therefore there are few ways in which episodes can be similar 
not just at the property level but at more abstract conceptual 
levels.\\ 

\subsection{Associativity is Detrimental for a 
Coarse-grained Memory}
The occasional episodic mind might (due to a 
genetic mutation) have a higher tendency toward associative thought, 
\textit{i.e.} a lower $\phi$, meaning that a recalled episode need 
not be very similar to the current episode to be evoked by it. Such 
an individual would be capable of making associations, perhaps even 
recursively, \textit{i.e.} engaging in a kind of representational 
redescription, or stream of self-triggered thought. However, recall 
from section \ref{sec:assocstructure} Langton's (1992) finding that 
the information-carrying capacity of a system is maximized when its 
interconnectedness or self-similarity of components falls within a 
narrow regime between order and chaos. Forcing this kind of 
interaction between concepts in a coarse-grained memory (one with low 
$n$) would cause discontinuity between one thought and the next. 
Therefore thought would readily become chaotic, lacking the necessary 
continuity to hone a solution to a problem or develop and carry out a 
plan. Abstract thought, unlike episodic thought, cannot rely on the 
continuity of the external world (\textit{e.g.} if a desk is in front 
of you now it is likely to still be in front of you now) to lend 
coherence to conscious experience; this continuity must be internally 
driven. Since when $n$ is low, a mutation that results in a high 
$\phi$ leads to discontinuous (garbled) thought, it would be 
evolutionarily selected against. \\ 

\subsection{Coarse-grained 
Memory Lacks Quantum Structure}
In summary, since the episodic mind 
contains only coarse-grained episodes and few concepts, there is 
little opportunity for interaction and transformation of conceptual 
structure through reminding events, abstraction, or concept 
combination. So for any $u \in {\cal U}$, represented by the SCOP 
$(\Sigma_u, {\cal M}_u, {\cal L}_u, \mu_u, \nu_u)$, any $p 
\in\Sigma$, for any context ${e, f, g...} \in {\cal M}$, $\mu(p, e, 
p)$ is close to an eigenstate, hence $\approx 1$. One does not need 
the formalisms of quantum mechanics to model the dynamics of the 
episodic mind precisely because concepts are few, and have little 
effect on one another. (Note that SCOP can still be used to describe 
the episodic mind because as mentioned earlier it can describe both 
classical and quantum dynamics.) 

\section{The Transition from 
Episodic to Mimetic Mind}
We noted that around 1.7 MYA there was a 
sudden increase in brain size, and this period is also associated 
with the origin of culture, and onset of what is referred to as the 
mimetic mind (Donald, 1991). The mimetic mind can access memories 
independent of external cues and act them out, group episodes 
together as instances of an abstract concept, refine ideas, and 
improve skills through repetition or rehearsal. We now proceed to 
examine a tentative scenario for how this transition came 
about.

\subsection{Associativity is Beneficial for a Fine-grained 
Memory}
It seems reasonable to assume that the larger brain of the 
mimetic mind has more neurons than the brain of the episodic mind, 
and thus responds to more properties or subsymbolic microfeatures of 
episodes. Therefore episodes are encoded in more detail, 
\textit{i.e.} $n$ is larger. There are more ways to distinguish one 
episode or item encoded in memory from another, and more pathways by 
which one episode can evoke representational redescription or a 
reminding of another episode. Thus it is well possible that for any 
given context ${e, f, g...} \in {\cal M}$, any $p \in \Sigma$ is not 
an eigenstate, i.e., $\mu(p, e, p) << 1$.\\
\indent Let us now examine 
what would happen if, again, due to a random mutation,  a larger 
brained mimetic individual had a heightened tendency toward 
associative thought, i.e., a high $\phi$. As with the episodic mind, 
this means that a recalled episode need not be very similar to the 
current episode to be evoked by it, and the individual is capable of 
representational redescription, or self-triggered thought. One 
thought is able to spontaneously evoke another thought that is a 
modified version of the first, and so forth recursively, such that 
the original thought transforms through interaction with the 
individual's conceptual structure. However, since memory is not 
coarse but fine-grained (large $n$), the outcome is different. The 
Gaussian distribution of possible Hamming distances between the 
current episode and an item encoded in memory (as depicted in Figure 
2) is wider, meaning there are now many more ways for an item in 
memory to be somewhat similar or loosely correlated to a given 
episode. This means that a stream of thought can retain the necessary 
balance between change and continuity to hone a solution to a problem 
or develop and carry out a plan. Since when $n$ is large a mutation 
that decreases $\phi$ thereby increasing $\mu$ leads to enhanced 
capacity for fruitful abstract thought, it facilitates a wide range 
of cognitive skills, and would be evolutionarily selected for.\\ 
\indent Because the individual with a higher $\phi$ is more prone to 
associative recall, this individual is better equipped to form 
concepts. Since there is more conceptual structure in place, there is 
more of it to be affected by any given episode, and it is more likely 
that conceptual structure thereby contributes to the context that 
evokes a successive change of state (self-triggered recall).\\

\subsection{Onset of Extensive Abstraction}
\indent The next step 
toward an integrated worldview is the formation of abstract concepts, 
such that items that share properties become conceived of as 
instances of the same thing. Suppose there are two episodes or 
concepts $u, v \in {\cal U}$, both modeled respectively by SCOPs 
$(\Sigma_u, {\cal M}_u, {\cal L}_u, \mu_u, \nu_u)$ and $(\Sigma_v, 
{\cal M}_v, \mu_v, \nu_v)$. Suppose there exists a concept $w \in 
{\cal U}$, such that $u$ and $v$ are exemplars (instances) of $w$, 
and hence can be represented as states of $w$. Suppose that concept 
$w$ is modeled by a SCOP $(\Sigma_w, {\cal M}_w, {\cal L}_w, \mu_w, 
\nu_w)$. Hence there are states $p, q \in \Sigma_w$, such that 
concepts $u$ and $v$ are states of $w$. We say that there exists a 
pathway from concept $u$ to concept $v$ and a pathway from concept 
$v$ to concept $u$, and denote these pathways $u \rightarrow v$ and 
$v \rightarrow u$. This can be generalized by saying that states 
$p_u$ of $u$ and $p_v$ of $v$ are also states of $w$.\\
\indent Let us now examine how conceptual structure forms, starting with basic 
level concepts and moving toward concepts that are more abstract 
(Rosch, 1978) \textit{via} the incorporation of new states and new 
contexts. Initially each concept has so few states associated with it 
that a state of one concept is rarely also a state of another 
concept. Thus reminding events are rare. However the activation of 
states of one concept that are also states of another is facilitated 
by the fact that the properties of any concept can themselves become 
viewed in conceptual terms, \textit{i.e.} they can become concepts. 
Consider the concept MONKEY. A mimetic individual might encode in 
memory a state $p$ of MONKEY with the property $a$: `has arms', 
\textit{i.e.} the state MONKEY WITH ARMS. If the individual has not 
just recognized that  `has arms' is a property of MONKEY but realized 
ARMS as a concept in its own right, then through the encoding of $p$, 
MONKEY WITH ARMS, the concept MONKEY becomes connected to the concept 
ARMS. The conceptualization of an episode as MONKEY WITH ARMS 
provides a pathway from episodes related to monkeys to episodes 
% L moved this from section 3 to here
related to arms.  For the combination MONKEY ARM, the concept MONKEY functions as a context rather than as an entity. This is clear because if we consider the states that MONKEY ARM can be in, these states are not states of MONKEY, but they are states of ARM. (Clearly MONKEY ARM is not a MONKEY, but MONKEY ARM is an ARM.) To model this, we let ${\cal H}$ be the Hilbert space of states of ARM and $|u\rangle$ be the ground state of ARM. Then MONKEY is described by an orthogonal projection operator $P: {\cal H} \rightarrow {\cal H}$, which is a linear function, which is idempotent, i.e., $P^2=P$, and Hermitian, i.e., $P^*=P$, such that the state of MONKEY ARM is given by ${P|u\rangle/\|P|u\rangle\|}$ where $\|P|u\rangle\|$ is the length of $P|u\rangle$. (It is possible to imagine a context where it would be better to consider MONKEY as the entity. Of course, knowing the rules of the English language, the combination should then be written as ARM MONKEY, and it would mean a type of monkey characterized for example by a special ways of using its arms. It is a characteristic of metaphor that contexts are pushed into becoming entities.)\\
% L moved this from section 3 to here
\indent Our mimetic individual might subsequently encode a 
state of MONKEY with the property $b$:  `has arms and legs'. The 
concept MONKEY is now connected to both concepts ARMS and LEGS. 
Moreover, `has arms' and `has arms and legs' are connected by the 
relation $b \le a$. Thus an early step in the forging of 
organizational structure amongst the items encoded in memory is the 
grouping together of those properties that exhibit spatiotemporal 
contingency. When the properties `has arms' and `has legs' co-occur, 
they start to be conceived of as parts of a whole. This organizing of 
stimuli into sets of properties that tend to go together in the world 
is an important step toward the establishment of conceptual structure 
that internally models that world. As episodes accumulate in memory, 
so do states and their SCOP structures. As memory comes to encompass 
more of the natural relations amongst concepts, connections amongst 
concepts become more widespread, and an integrated conceptual 
structure starts to emerge. \\
\indent The same thing happens with 
contexts (as one might expect, since as stated earlier concepts can 
also be contexts). Consider the concept STAR. One might encode in 
memory a state of STAR under the context $e$:  {\it to the north}. 
The concept STAR is thereby connected to the concept NORTH. One might 
subsequently encode a state of STAR under the context $f$:  {\it to 
the north at dusk}. The concept STAR is now connected to the concept 
DUSK as well as the concept NORTH, and the contexts $e$:  {\it to the 
north} and $f$: {\it to the north at dusk} are connected by the 
relation $f \le e$. As memory comes to encompass contexts that are 
increasingly stronger, connections amongst concepts become more 
widespread, and the worldview thereby becomes more 
integrated.\\
\indent As concepts accumulate, episodes are 
increasingly interpreted in terms of them, making episodes 
increasingly likely to evoke reminding of previously encoded 
episodes, or bring about representational redescription of the 
current episode. The probability increases that a given episode is 
similar enough to at least one item in memory to evoke it. With SCOP 
this is modeled as an increase in the number of potential contexts 
for which $\mu(p_i, e_j, p_k) > 0$. The upshot is that one is able to 
recursively modify and redescribe thoughts and events. As more 
concepts become available, this gives rise to self-triggered 
sequences of thought, which increase in duration and 
frequency.\\
\indent The probability of one MONKEY episode evoking 
another is enhanced through the recognition that they are in essence 
the same thing, \textit{i.e.} instances of the concept MONKEY. The 
concept MONKEY facilitates the evoking of memories involving 
instances of monkeys. It connects any given monkey episode with not 
just other instances involving monkeys but with whatever associations 
have been formed amongst the concept MONKEY and other items (such as 
ARMS, LEGS...). Since representational redescription is now possible, 
using SCOP the function $\mu(q, e, p)$ gives a probability greater 
than zero that a particular state under a particular context will 
change to a particular other state, \textit{e.g.} LARGE MONKEY might 
transform to DANGEROUS LARGE MONKEY.\\
\indent A new state of a 
concept constitutes a new episodes, hence with each newly-formed 
state of a concept the number of episodes increases. If it is an 
episode that does not map onto a particular thing or event in the 
world but an abstraction (such as FURNITURE, or BEAUTY, or 
DEMOCRACY), it increases not just the number of items amongst which 
one could form associations, but the density of connections amongst 
previously encoded items. For example, due to a memory of someone 
using twine to strangle something, TWINE may be associatively linked 
to ARROW though they share few properties, because they are both 
instances of WEAPON. Depending on the context, a concept such as 
WEAPON can project onto states with different properties and 
different meanings. It is because there is greater potential in the 
mimetic mind for interaction amongst episodes (both conceptualized 
and unconceptualized) that a Hilbert space quantum model is well 
suited to model its dynamics.

\subsection{Strengths and Limitations 
of the Mimetic Mind}
It is tentatively proposed that this state of 
affairs describes a mimetic mode of cognition.  It is not merely a 
relay-station for coordinating inputs to appropriate responses. 
Because $n$ is higher, $\phi$ can afford to be lower, which 
facilitates the forging of meaningful conceptual structure. The 
mimetic mind engages extensively in abstraction and is capable of 
self-triggered recall. Since memories and concepts are starting to be 
organized in a way that reflects abstract relationships amongst them, 
reminding events are not infrequent, and potentially chained to give 
a short stream of self-triggered thought, and the current thought can 
become part of the context in which the next is experienced; thus 
actions can be refined. However, the mimetic mind is not very 
creative, and it is not integrated. it does not constitute a coherent 
mental model of how various aspects of the world relate to one 
another. \\

\section{The Transition from Mimetic Mind to Modern 
Human Mind}
The archaeological record reveals a burst of creativity 
indicative of a cultural transition in Europe between 60,000 and 
30,000 ka (Bar-Yosef, 1994; Klein, 1989; Leakey, 1984; Mellars, 1973, 
1989a, b; Mithen, 1996, 1998; Soffer, 1994; Stringer \& Gamble, 1993; 
White, 1982). More recently it has been suggested that modern 
cognition originated earlier and more gradually in Africa, and that 
the European data reflects a wave of immigration of humans from 
Africa with sufficient intelligence and creative problem solving 
skills to cope with the new environment (McBrearty \& Brooks, 2000). 
Leakey (1984) writes ``Known as the Upper Paleolithic Revolution, 
this collective archaeological signal is unmistakable evidence of the 
modern human mind at work" (p. 93-94). Mithen (1996) refers to this 
period as the `big bang' of human culture, claiming that it marks the 
beginning of art, science, and religion, and that it shows more 
innovation than the previous six million years of human evoluti
 on. 
What brought about this cultural transition? \\
\indent One 
possibility is that the cultural transition of the Middle/Upper 
Paleolithic reflects an enhanced ability to blend concepts 
(Fauconnier \& Turner, 2002). In a similar vein, Mithen (1998) 
suggests that it reflects onset of the capacity to transform 
conceptual spaces. The answer proposed here is consistent with these 
hypotheses but further specifies what \textit{kind} of cognitive 
structure could have arisen to allow for these 
abilities.

\subsection{Onset of the Capacity to Shift between 
Analytic and Associative Thought}
It is suggested that the transition 
from a mimetic to a modern form of cognition is what gave rise to the 
cultural transition of the Middle/Upper Paleolithic, and that this 
came about through onset of the capacity to subject different 
situations to different kinds of cognitive processing. It is widely 
believed that there are two forms of thought, or that thought varies 
along a continuum between two extremes (Ashby \& Ell, 2002; Dennett, 
1978; Gabora, 2002a, b; Gabora, under revision; Johnson-Laird, 1983; Neisser, 
1963, Piaget, 1926; Rips, 2001; Sloman, 1996). At one end of the 
continuum is an \textit{analytic} mode of thought conducive to 
deduction and to describing and analyzing relationships of cause and 
effect. At the other end of the continuum is an intuitive, 
overinclusive, or \textit{associative} mode of thought conducive to 
finding subtle relationships; \textit{i.e.} connections between items 
that are correlated, but not necessarily causally related. In this 
mode, items are represented in terms of not just their most typical 
properties but also in terms of less typical, perhaps 
context-dependent ones as well. Associative thought is related to the 
notion of flat associative hierarchies, a term applied to those who 
give not only typical but also marginal or atypical items when asked 
to say words that come to mind in response to a particular word, 
\textit{e.g.} OSTRICH in response to BIRD, or BEANBAG CHAIR in 
response to CHAIR (Mednick, 1962). The notion of two forms of thought 
is also captured in G\"ardenfors' tiered model of cognition; the 
analytic mode is related to the symbolic level of cognition and the 
associative mode is related to the conceptual level. \\
\indent It 
has been proposed that the transition from mimetic mind to modern 
human mind came about through the onset of \textit{contextual focus}: 
the capacity to spontaneously  focus or defocus attention in response 
to the current situation, and thus alternate between analytic and 
associative thought according to the demands of the situation 
(Gabora, 2003, Aerts \& Gabora, 2005a). This allowed for a worldview 
with recursively embedded hierarchical structure and concepts of 
varying levels of abstraction, which paved the way for conceptual 
integration. Let us now formulate this hypothesis more 
precisely.\\
\indent Here we describe the onset of the capacity for 
contextual focus mathematically as the onset of the capacity to 
modulate $\phi$. It is proposed that analytic thought is the result 
of focusing attention, thereby increasing $\phi$, which globally (for 
all states of all concepts, as well as all unconceptualized cognitive 
states) decreases $\mu$, the transition probability. Since one state 
is less likely to spontaneously give way to another as a consequence 
of their shared properties and abstract conceptual structure, effort 
can be focused on manipulating items in their most prototypical, 
unambiguous form according to the rules of logic, as in the formation 
of abstract concepts, the recognition of episodes as instances of a 
concept, and furthering development of the structure of the lattices 
of properties and contexts associated with known, existing concepts 
as described above. \\
\indent It is proposed that associative 
thought is the result of defocusing attention, thereby decreasing 
$\phi$. One is more likely to make associations and be reminded of 
items based on not just their most typical properties but also in 
terms of less typical, context-dependent properties, or on the basis 
of abstract structure, or spatiotemporal contiguity. This is 
conducive to applying known concepts in new contexts; for example, 
application of the concept HANDLE to challenging situations, as in 
`get a handle on the situation' or to intimate relationships, as in 
`love handle'. This results in a transition from domain-specific 
representations to more broadly applicable representations, enabling 
seemingly disparate domains of knowledge to be bridged.

\subsection{The Formation of Conjunctions}
% LG ch
The capacity to 
temporarily decrease $\phi$ and thereby enter a more chaotic regime 
(for example, when one is stuck in a problem or seeking 
self-expression) is not just conducive to applying concepts in new 
contexts, but also to merging concepts to generate new ones. This is 
described as follows. \\
\indent For two concepts $u, v \in {\cal U}$ 
described by SCOPs $(\Sigma_u, {\cal M}_u, {\cal L}_u, \mu_u, \nu_u)$ 
and $(\Sigma_v, {\cal M}_v,$ $\mu_v, \nu_v)$ it may be that a 
conjunction exists. A conjunction is described in SCOP using the 
tensor product when the SCOPs are Hilbert space based, as described 
in section \ref{sec:multipleconceptsHibert}. The conjunction is 
considered a new concept, not just the logical construct of a 
conjunction. For example, suppose $u$ is the concept SNOW and $v$ the 
concept MAN. The tensor product SCOP models the concept SNOWMAN. 
States of SNOWMAN are states of SNOW and states of MAN. For example 
OG WITH SNOW ON HIS HEAD, as one concept, is not a state of the 
concept SNOWMAN, but it is a state of the logical construct of the 
conjunction SNOW and MAN.\\ 
\indent The formation of a conjunction 
involves two steps (Aerts \& Gabora, 2005b). The first step involves 
recognition that one ofthe concepts is a specific kind of context for 
the other. For example, consider the concepts SNOW and MAN. Note that 
{\it The man is made of snow} can be a context for MAN. It changes 
the ground state of MAN to a state of MAN that is a `prelude' to the 
formation of the new concept SNOWMAN. This prelude state is the root 
of the pathway that connects the concepts SNOW and MAN.\\
\indent We 
define this pathway formally as follows. Consider two concepts $u, v 
\in {\cal U}$, modeled by SCOPs $(\Sigma_u, {\cal M}_u, {\cal L}_u, 
\mu_u, \nu_u)$ and $(\Sigma_v, {\cal M}_v, \mu_v, \nu_v)$. Supposes 
that state $p \in \Sigma_u$ (hence a state of concept $u$) is a 
context of concept $v$, which we call $e_p \in {\cal M}_v$. This 
context $e_p$ creates a new state of concept $v$ by acting on the 
ground state $\hat q$ of $v$. We could for example denote this new 
state as $\hat p \triangle q$. We say that a pathway exists from 
$\hat p$ to $q$, denoted $\hat p \leftrightarrow q$. State $\hat p 
\triangle q$ is a prelude to the state of the new concept when a 
conjunction of $u$ and $v$ forms. For example, the state of MAN that 
results when the context {\it the man is made of snow} influences the 
ground state of MAN is the state $\hat p \triangle q$. It is what we 
are calling a `prelude state' of the ground state of the new concept 
SNOWMAN.\\

\subsection{Conceptual Integration}
\indent The formation 
of a new concept increases the number of concepts by one, but it 
increases the number of pathways by, not just the number of states of 
that concept, but also other episodes and concepts that share 
conceptual structure (\textit{e.g.} BEANBAG CHAIR is a state of 
BEANBAG and a state of CHAIR and for that matter a state of 
FURNITURE\dots). Therefore, as the number of concepts increases, the 
number of pathways  increases faster. It is known in graph theory 
that when the ratio of edges to points reaches approximately 
0.5---sometimes referred to as the \textit{percolation threshold} 
(Kauffman, 1993)---the probability that one giant cluster emerges 
goes from extremely unlikely to almost inevitable (Erdos \& Renyi, 
1959, 1960). Here, when the ratio of pathways to episodes and 
concepts surpasses this threshold, it becomes almost inevitable that 
the items in memory reach a critical density where a large subset of 
them undergoes a phase transition to a state wherein  for each 
episode or concept there exists \textit{some} possible direct or 
indirect route to any other episode or concept by way of internally 
driven remindings or associations. In this way, they constitute an 
integrated conceptual web. \\
\indent There are numerous ways of 
achieving conceptual integration even starting from the same set of 
episodes, and these ways convey varying degrees of integration and 
consistency on the resulting worldview. They differ with respect to 
the lattice structures of states and contexts that connect these 
episodes, and the degree of penetration of the conceptual space. 
Under what conditions will the transformation from discrete memories 
to interconnected conceptual web actually occur? We need to show that 
some subset of the concepts encoded in an individual's mind 
inevitably reach a critical point where there exists a direct or 
indirect route (sequentially applied pathways) by which every member 
of that subset can access every other member. We need to show that 
$R$, the number of pathways by which one episode can evoke another, 
increases faster than $s$, the number of episodes. That is, as the 
memory assimilates episodes, it comes to have more ways of generating 
them than the number of them that has explicitly been 
encoded.\\
\indent Under what conditions does $R$ increase faster 
than $s$? The ability to form context-sensitive concepts plays a 
crucial role. To determine how this affects  $R$, let us assume for 
the moment that memory is fully connected. In other words, we 
temporarily assume that all states of all encoded items have a 
non-zero transition probability to all states of all other encoded 
items. Clearly this is not the case, but this simplification facilitates 
analysis that also applies to a more realistic model of memory. To 
make things simple we will be conservative and limit the sort of 
pathways under consideration to associative recall and pathways 
between abstract concepts and their instances. Concepts have $n$ 
properties, where the number of applicable properties (properties 
with non-zero weights) ranges from a minimum of $m$ to a maximum of 
$M$. $R_A$, the number of ways a pathway can be realized through 
abstraction, equals the number of pathways enabled by an 
$n$-dimensional concept, multiplied by the number of n-dimensional 
concepts, summed over all values of $n$ from $m$ to $M-1$. The number 
of pathways equals the number of items that are instances of an 
$n$-dimensional concept $= 2^{M - n}$. The number of $n$-dimensional 
concepts is equal to the binomial coefficient of $M$ and $n$. The 
result is multiplied by two since an a concept can evoke an instance, 
and likewise, an instance can evoke a concept.
\begin{eqnarray}
	R_A = 2(2^{M-m}{M \choose m} + 2^{M-(m+1)}{M \choose m + 1} + 
... + 2{M \choose M - 1}) \\
	= 2\displaystyle\sum_{n=m}^{M-1} 
2^{M-n}(n-1){M \choose n}
\end{eqnarray}\\
\indent This shows that 
the greater the density of concepts, the greater the likelihood of 
conceptual integration. Abstraction (the formation of a general 
concept out of instances) increases $s$ by creating a new concept. 
But it increases $R$ more, because the more abstract the concept, the 
greater the number of concepts a short Hamming distance away (since 
$|x_i -k_i| = 0$ for the irrelevant properties, so they make no 
contribution to Hamming distance). In other words, concepts with 
fewer properties enable exponentially more pathways.\\
\indent  A 
second thing to note is that as $n$ starts to decrease, the number of 
possible abstractions for each value of $n$ increases (up to $M/2$ 
after which it starts to decrease). Thus the more likely any given 
memory item is to get activated and participate in a reminding or 
associative recall event. Whereas $R$  increases as abstraction makes 
relationships amongst concepts increasingly explicit, $s$ levels off 
as new experiences have to be increasingly unusual in order to result 
in the formation of new concepts. Furthermore, when the carrying 
capacity of the memory is reached (such that the encoding of new 
items entails significant degradation of previously encoded ones), 
$s$ plateaus but $R$ does not. Taken together these points mean: the 
more deeply a mind delves into abstractions, the more the 
distribution in Figure 2 rises and becomes skewed to the left. The 
effect is magnified by the fact that the more active a region of 
conceptual space, the more likely an abstraction is to be positioned 
there, such that abstractions beget abstractions through positive 
feedback loops.\\
\indent  Thus, as long as $\phi$ is low enough to 
permit abstraction and small enough to permit temporal continuity, 
the average value of $n$ decreases, and sooner or later, the system 
is expected to reach a critical percolation threshold such that $R$ 
increases exponentially faster than $s$. The memory becomes so 
densely packed that any episode is bound to be close enough in 
Hamming distance to some previously-encoded episode to evoke it. The 
memory (or some portion of it) is holograph-like in the sense that 
there is a pathway of associations from any one episode to any other. 
What was once just a collection of isolated memories is now a 
structured network of concepts, instances, and relationships---a 
worldview. It can engage in streams of thought that redefine elements 
of the world in terms of their substitutable and complementary 
relationships, plan a course of action, engage in symbolic 
communication with others, and so forth, as outlined in the 
introduction. It is used not just passively to understand what the 
world is like, but actively to navigate life experiences. 
\\

\subsection{Conceptual Integration in SCOP}
Let us now show what 
conceptual integration means with respect to SCOP. For an arbitrary 
subset $X \subset {\cal X}$ of episodes we define
\begin{equation}
cl(X)=\{y\ 
\vert y \in {\cal X},{\rm and}\ \exists \ x \in {\cal X}\ {\rm such\ 
that}\ y \leftrightarrow x\}
\end{equation}
% tried moving to appendix
In Appendix 1 we prove that $cl$ is 
a closure relation. This allows us to consider for an arbitrary 
subset of episodes $X \subset {\cal X}$ its closure $cl(X)$. This is 
the set of episodes such that for each episode $x \in cl(X)$ there is 
an episode $y \in X$ such that $x \leftrightarrow y$ is connected to 
$y$.\\
 \indent We now consider `one' specific episode $x$, and the subset 
containing `only' this one episode, hence the singleton $\{x\}$. We 
can ask what is the closure of this subset, hence what is 
$cl(\{x\})$? In other words, what is the set of episodes connected to 
$x$? In Appendix 2 we show that the closure relation conserves full connectedness. This means 
that if we start by considering a fully connected subset of episodes, 
and add all the episodes connected to episodes of this subset, this 
is still a fully connected subset of episodes.\\
 \indent The next thing we want to show is that $cl(X)$ is the biggest fully 
connected subset containing $X$. This proof is given in Appendix 3. Let us investigate 
what this tells us about the set of all episodes ${\cal X}$. Consider 
two episodes $x, y \in {\cal X}$. There are theoretically only two 
possibilities. Either $x \leftrightarrow y$, or $x 
\not\leftrightarrow y$. If $x \leftrightarrow y$, we can consider 
$cl(\{x,y\})$, which is then the biggest fully connected subset 
containing $x$ and $y$. If $x \not\leftrightarrow y$ we consider 
$cl(\{x\})$ and $cl(\{y\})$. We know that $cl(\{x\})$ is the biggest 
fully connected subset containing $x$ and $cl(\{y\})$ is the biggest 
fully connected subset containing $y$. Can $cl(\{x\})$ and 
$cl(\{y\})$ have elements in common? No, indeed, suppose that $z \in 
cl(\{x\}) \cap  cl(\{y\})$. Then $z \leftrightarrow x$ and $z 
\leftrightarrow y$, and from transitivity of the connectivity 
relation it follows that $x \leftrightarrow y$. Hence $cl(\{x\}) \cap 
cl(\{y\})=\emptyset$. But the situation is more extreme. We do not 
even have any element in $cl(\{x\})$ that is connected with any 
element in $cl(\{y\})$. We can prove the 
proposition that for 
the closure of two single episodes, the closure subsets are equal, or 
their intersection is empty. This proof is given in Appendix 4. In Appendix 5 we prove that ${\cal X}$ is equal to the disjoint union of fully connected 
subsets of episodes, and moreover each of these 
fully connected subsets of episodes is the closure of a single 
episode. This implies that if one of these regions of fully connected 
subsets becomes dominant, in the sense that it is the one used to 
generate internal contexts and interpret episodes, and the one 
generally activated in conscious experience, then also this one is 
the closure of one single episode. This one single episode is not 
unique; there are many episodes that `as one single episode' can play 
the role of being the generating element for the whole closure. This 
is the meaning of the possibility that $cl(\{y\})=cl(\{z\})$ is many 
occasions, and in fact whenever $y \leftrightarrow z$. Others may 
continue to exist but be virtually inaccessible from the dominant 
one, and may be experienced only in situations where there are fewer 
social and pragmatic constraints on how we internally generate the 
contexts  that enable thoughts to unfold. The bottom line is that the 
conceptual structure now has a vast amount of potentiality, and the 
extent to which this potentiality is accessed at any given instant 
depends on the situation and the value of $\phi$.\\

\section{The 
Development and Evolution of an Integrated Worldview}
\indent It is 
possible that some items in memory are isolated outliers---not woven 
into this conceptual web, and thus inaccessible. (It seems reasonable 
that this might be the case for experiences that are far from 
everyday reality and insufficiently reinforced through reminding 
events, such as dreams or perhaps traumatic experiences.) Since we do 
not have conscious access to them, it is difficult to know whether 
they are abundant or rare. It is possible that much of what 
constitutes a mind is never woven into an integrated worldview. What 
we can say is that to constitute a worldview, the memory architecture 
must be such that, for any typically encountered stimulus or 
situation, there must exist a means of relating or re-describing it 
in terms of previously encoded items. Thus for an integrated 
worldview, this includes all memory items necessary for the 
redescription and accommodation of frequently encountered 
experiences. \\
\indent Note that conceptual integration would be 
unlikely to be achieved simultaneously throughout a society of 
individuals. It would thus take time for social propagation of the 
cultural `fruits' of such an achievement. However, even if for 
example one individual, Agu has a higher activation threshold than 
another individual, say Oga, once he has assimilated enough of Oga's 
abstractions, his concepts become so densely packed that a version of 
Oga's worldview snaps into place in his mind. Agu in turn shares 
fragments of his worldview with his friends, who in turn shares them 
with others. These different `hosts' expose their versions of these 
fragments of what was originally Oga's worldview to different 
experiences, different bodily constraints, sculpting them into unique 
internal models of the world. Small differences are amplified through 
positive feedback, transforming the space of viable worldview niches. 
Individuals whose activation threshold is too small to achieve 
worldview closure are at a reproductive disadvantage and, over time, 
eliminated from the population. When Agu imitates a mannerism of 
Ogu's he might, on the fly, modify it by putting his own spin on it. 
As worldviews become increasingly complex, the artifacts they 
manifest in the world become increasingly complex, which necessitates 
even more complex worldviews, and so forth. Thus a positive feedback 
cycle sets in. The elements of human culture---the artifacts, the 
stories, the languages in which they are told---reflect the states of 
the worldviews that generated them, specifically the associative 
structure amongst concepts (and the words by which they are 
accessed), which in turn reflects their underlying motives and 
desires. \\
\indent This analysis suggests that animals are not 
prohibited from evolving complex cognition a priori, but that without 
the physical capacity to generate and manipulate complex artifacts 
and vocalizations there is insufficient evolutionary pressure  to 
tinker with $\phi$, the association threshold, until it achieves the 
requisite delicate balance to sustain a stream of thought, or to 
establish and refine the necessary feedback mechanisms to dynamically 
tune it to match to the degree of associativity or `conceptual 
fluidity' required at any given instant. It may be that humans are 
the only species for which the benefits of this tinkering process 
have outweighed the risks. This would make sense because with the 
capacity for complex speech and hand movement there are clearly more 
ways to manifest complex thoughts, i.e., more ways in which they could 
yield outcomes in the world.\\
\indent It was noted that it is not 
the \textit{presence of} but the \textit{capacity for} an integrated 
worldview that the human species has come to possess. Following the 
pioneering efforts of Piaget, Vygotsky and more recently others, it 
has become clear that a worldview is not present from birth but 
develops naturally through experience in the world. The infant mind 
is predisposed to selectively attend to biologically salient 
features, and respond accordingly. If it is hungry and sees its 
mother's breast, it suckles; if it feels something extremely hot or 
cold it pulls away, and so forth.  In addition to innate 
predispositions to respond categorically to certain stimuli, it is 
widely thought that infants possess higher cognitive competencies 
(Gelman, 1993; Keil, 1995). These competencies may be due to core 
knowledge (Spelke, 2000), intuitive theories (Carey, 1985), or simply 
predispositions to direct attention to salient (particularly social) 
elements of a situation (Leslie, 2000). An infant is also capable of 
storing episodes as memories. Although episodes from infancy are 
rarely accessible later in life, they \textit{do} get etched into 
memory, as evidenced by the capacity for reminding events, which is 
present by two month of age (Davis \& Rovee-Collier, 1983; 
Rovee-Collier \textit{et al.}, 1999; Matzel \textit{et al.}, 1992) 
and possibly earlier (Rovee-Collier, pers. comm). Thus, an infant is 
born predisposed toward conceptual integration, but the process must 
begin anew in each young mind. The infant may be born with innate 
concepts, but its memory is not densely packed enough (i.e., 
distributed representations do not overlap enough) to prompt 
recursive redescription or reminding events. Even if items are 
encoded in detail, that is, even if \textit{n}, the mean number of 
properties per encoded item, is large, there are too few of them for 
reminding events to occur with high frequency. Returning to Figure 2, 
there are few items within a given Hamming distance of any item. 
Moreover, it lacks the conceptual structure that facilitates efforts 
to understand and make sense of how episodes and their contents are 
related to one another.\\
\indent Although the issue is 
controversial, it is widely accepted that between six and eight years 
of age, a child moves from implicit, domain-specific representations 
to explicit, more broadly applicable representations 
(Karmiloff-Smith, 1990, 1992). Aided by social exchange, and mediated 
by artifacts, a framework for how things are and how things work 
falls into place in a child's mind, and it bears some likeness (and 
also some dis-similarities) to that of its predecessors, such as the 
worldviews of parents and other influential individuals. Some 
experiences are either so consistent, or so inconsistent with the 
child's worldview that they have little impact on it. Others mesh 
readily with existing ideas, or ring true intuitively, and percolate 
deep into the worldview, renewing the child's understanding of a 
myriad other notions or events. The child is thereby encultured, and 
becomes a unique cog in the culture-evolving machinery. Human life 
offers a continuous stream of situations that disrupt the state of 
conceptual integration. For example, problems arise that cannot be 
consistently explained in terms of other elements of the worldview as 
they are currently understood. The individual may reconceptualize the 
problem or situation from the perspective of a context generated by 
the worldview. The process is recursive in that the new cognitive 
state is a function of both the previous cognitive state and the 
context. Each internally generated context may induce a change in the 
conception of the problem or situation, and each change in conception 
of the problem may modify he worldview, until it eventually recoups 
the state of conceptual integration.\\
\indent The idea of a 
conceptually integrated worldview sheds light on the thorny issue of 
the role of evolution in cognition. Hampton (2004) argues that if 
evolutionary psychology as a form of strong adaptationism insists 
that novel behaviors are the product of directly selected 
adaptations, it follows that there can be no evolutionary account of 
many modern activities such as driving and reading. However, one of 
the four approaches subsumed under by evolutionary psychology, as 
laid out in \textit{The Adapted Mind} (Barkow \textit{et al.} 1993), 
is the study of culture as an evolutionary process in its own right. 
Even if modern activities defy explanation in terms of biological 
evolution they may be explicable in terms of cultural evolution. As 
Deacon (1997) points out, language evolution is probably thousands of 
times more rapid than brain evolution. Bouissac (1998) concludes from 
this that ``such a parasitic theory of language evolution is rooted 
in Dawkins' meme hypothesis". However, the notion of conceptual 
integration suggests an alternative: that it is not discrete memes 
(transmittable ideas or artifacts) that evolve through culture, but 
conceptually integrated networks of them---worldviews. The argument 
for this is laid out in detail elsewhere (Gabora 2004); here it is 
summarized insofar as is necessary and relevant.\\
\indent A basic 
tenet of Dawkins' proposal is that the meme is replicator---an entity 
that makes copies of itself. Clearly ideas and artifacts are not 
replicators of the same sort as modern day organisms, which use a 
DNA- or RNA- based \textit{self-assembly code}. A telltale sign of 
self-assembly coded replication is restricted inheritance of acquired 
characteristics. The effects of contextual interactions between the 
organism and its environment are wiped out in the next generation, 
and do not affect the evolution of lineages. Thus for example, if one 
cuts off the tail of a mouse, its offspring have tails of normal 
length. As many have noted, however, acquired characteristics may be 
inherited in cultural transmission. If you hear a joke and then put 
your own slant on it in the retelling of it, those who pass it on 
subsequently will likely pass on your changed version.) An idea may 
\textit{retain} structure as it passes from one individual to another 
(like a radio signal intercepted by a new radio), but does not 
\textit{replicate} it.\\
\indent Cultural artifacts are not the only structures to have evolved without a self-assembly code. Our 
application of closure to the origin and evolution of an integrated 
worldview was inspired by current thinking on  the problem of how 
life began (Gabora, 2006; Kaufman, 1993; Morowitz, 1992; Vetsigian, 
Woese, \& Goldenfeld, 2006; W?echtersh?euser, 1992; Weber, 1998, 
2000; Weber \& Depew, 1996; Williams \& Frausto da Silva, 1999, 2002, 
2003; Woese, 2002). For example, Kaufman (1993) uses the concept of 
\textit{autocatalysis} to describe how the earliest self-replicating 
structures could have emerged. The autocatalytic sets of polymers 
widely believed to be the earliest form of life generated 
self-similar structure, but since there was no genetic code yet to 
copy from, there was no explicit copying going on. The presence of a 
given catalytic polymer, say polymer X, simply sped up the rate at 
which certain reactions took place, while another catalytic polymer, 
say Y, influenced the reaction that generated X. Eventually, for each 
polymer in the set, there existed a reaction that catalyzed it. Thus 
the set as a whole was autocatalytic, able to replicate itself. 

\footnote[5]{Some authors  (e.g. Thompson, 2007) claim that life 
requires a semi-permeable boundary to allow for spatial demarcation 
and containment. We believe this is unnecessary because in an 
autocatalytic network, if there is a molecule that catalyzes a 
particular molecule X, then molecule X can effectively enter and 
become part of the network. If there is no molecule that catalyzes 
molecule X, then molecule X is effectively excluded. Possession of a 
membrane boundary thus strikes us as a redundant, arbitrary 
requirement, true of life as we know it but not life as it could be; 
one can imagine entities that are alive but nonetheless interwoven 
without clear-cut physical boundaries separating them. Nevertheless it is 
indeed the case that semipermeable membranes play an exceedingly 
important role in life and may turn out to be a requirement.} 
These 
earliest life forms have been shown to have evolved through a 
non-Darwinian process involving not natural selection operating at 
the population level but communal exchange of innovations operating 
at the individual level.\\
\indent Kauffman (1993) argues that it is 
extremely unlikely that the earliest life forms themselves replicated 
using a self-assembly code, though such a code could have arisen by 
way of a simpler, uncoded self-replicating structure. His proposal is 
that the first living organism was a set of autocatalytic polymers. 
Since there were no self-assembly instructions to copy from, there 
was no explicit copying going on. The presence of a given catalytic 
polymer, say X, simply speeded up the rate at which certain reactions 
took place, while another polymer, say Y, influenced the reaction 
that generated X. Eventually, for each polymer, there existed another 
that catalyzed its formation. Such autocatalytic sets have been 
referred to as \textit{primitive replicators} because they generate 
self-similar structure, but in a self-organized manner, 
through bottom-up interactions rather than a top-down self-assembly 
code (such as the genetic code), thus they replicate--or more precisely, regenerate themselves--with low 
fidelity (Gabora 2004, 2006, 2008). They are self-mending in the sense that if the web of catalytic reactions breaks down (if, for example, one of the polymers gets used up) one or more others is eventually found to take its place, such that the network of reactions becomes again able to regenerate itself. \\
\indent 
It may be that the emergence of this kind of  complex, adaptive self-modifying, self-mending, self-regenerating (i.e., autopoietic) structure is the critical step in the origin of any kind of evolutionary process,  be it biological, cultural, or some other sort that we may not have identified. The evolution of culture involves not just creativity, nor just imitation, but a proclivity to put our own spin on the inventions of others, to use them for our own purposes and adapt them to our own needs and desires, such that the process  becomes cumulative, adaptive, and open-ended. The present paper describes steps toward the kind of cognitive structure that is capable of supporting this process. Like Kauffman's (1993) autocatalytic sets of polymers, an integrated 
worldview is autopoietic. Just as polymers catalyze reactions 
that generate other polymers, the retrieval of an item from memory 
can in turn trigger other items, cross-linking memories, ideas, and 
concepts into an integrated conceptual structure. Moreover, this conceptual structure regenerates itself; it constitutes a primitive replicator. Its elements (ideas, stories, and so forth) evoke one 
another through deduction, association, and reminding events, 
frequently resulting in their expression in the form of cultural 
artifacts, or get communicated and incorporated into the worldviews 
of other individuals. 
The result is that contextual interactions can 
affect the `cultural lineage' downstream and thus play an important 
role in their evolution. Thus a worldview gets regenerated in a 
piecemeal fashion through social learning, both directly and mediated 
by artifacts. Ideas and artifacts are not full-fledged replicators 
themselves, but manifestations of a worldview, which is a primitive 
replicator. They participate in the evolution of culture by revealing 
certain aspects of the worldview that generated them, thereby 
affecting the worldviews of those exposed to them.\\
\indent In Kauffman's (1993) computer simulations of the origin of life through 
autocatalytic closure, each polymer was composed of up to a maximum 
of M monomers, and assigned a low \textit{a priori} random 
probability \textit{P} of catalyzing each reaction. The lower the 
value of $P$, the greater $M$ had to be, and \textit{vice versa} in 
order for autocatalytic closure to occur. Computer simulation of 
conceptual integration might be expected to reveal a similar 
trade-off between $M$, which in the cognitive scenario refers to the 
maximum number of dimensions along which episodes are encoded, and 
$\phi$, the probability for items in memory to evoke one another. We 
might also expect to see that, as with biological life, worldviews 
that replicate themselves more efficiently will proliferate at the 
expense of others that are less efficient. Indeed, we are compelled 
to engage in behavior that helps others absorb the knowledge, 
attitudes, and ways of doing things that constitute our worldview, 
and those that were not so inclined in the past may well have been 
selected against at this cultural level.

\section{Summary and Discussion}
Our capacity to adapt ideas to new situations, see one 
thing in terms of another, blend concepts together in an endless 
variety of ways to interpret and express real or imagined situations, 
are all indicative of the integrated nature of a human worldview. 
% change L
This paper has presented a speculative attempt to formalize the process by which conceptual integration comes about. The paper does not tell a finished story but merely lays some groundwork and points the way. Several elements will require further development, such as the use of the Schr\"odinger equation to  model the dynamical unfolding of the conception of combinations of concepts. \\
\indent Let us retrace the major steps taken.
% end change L
The integrated quality of a worldview---the capacity for interpenetration 
of one thought or idea into another---presents a chicken and egg 
paradox. Until memories are woven into an integrated worldview, how 
can they generate the reminding or evoking events that constitute a 
stream of thought? Conversely, until a mind can generate streams of 
thought, how does it integrate memories into a worldview? How could 
something composed of complex, mutually dependent parts come to be? 
Moreover given the chameleon-like behavior of concepts,  and what 
kind of mathematical description can we give to them? In this paper 
it is proposed that a worldview emerges through conceptual 
integration yielding a structure whose dynamics requires the quantum 
formalism for its description.\\
\indent Let us retrace the steps in 
how closure and SCOP are used to model the transition to an 
integrated worldview. The associative structure of the episodic minds 
of \textit{Homo habilis}, our earliest ancestors was coarse-grained; 
items were not encoded in detail. Thus we model the \textit{Homo 
habilis} mind as having little interaction amongst concepts, little 
opportunity for reminding events and associations, no pressing need 
for a quantum formalism, and no conceptually integrated worldview. 
\\
\indent The memory of the mimetic mind of \textit{Homo erectus} is 
more fine-grained; items are encoded in greater detail. Thus there is 
opportunity for reminding events and associations, and interaction 
amongst concepts, and a need for the quantum formalism to model how 
concepts emerge from a ground state to an actual state in a manner 
that depends on the constellation of other evoked concepts. There is 
concept formation through abstraction, modeled as the recognition 
that episodes or concepts are instances of a more general concept, 
and the accumulation of sets of states, contexts, properties, weights 
and transition probabilities. There is formation of conceptual 
structure modeled as the formation of lattices of properties and 
contexts through the realization of natural relations. However, 
because there is no ability to modulate the degree of interaction 
amongst concepts depending on what is called for by the situation, 
the conceptual web is crude and fragmented.\\
\indent The modern 
human mind has the ability to shift between analytic thought, 
conducive primarily to realizing relationships amongst states of a 
known concept, and associative thought, conducive primarily to 
forging new concepts through the formation of conjunctions, which are 
entangled states that result through application of the tensor 
product of the Hilbert spaces of the two constituent concepts. It is 
proposed that the penultimate step toward achieving an integrated 
worldview was to acquire the capacity to spontaneously focus 
attention (conducive to analytic thought) or defocus attention 
(conducive to associative thought) depending on the circumstance. 
This is modeled as onset of the modulation of $\mu$, the transition 
probabilities using a variable we called $\phi$. Once the capacity 
has evolved to alter $\phi$ according to the situation, analytic 
thought and associative thought can work in concert to organize and 
reorganize conceptual structure. Analytic thought enables the 
identification of causal relationships, while associative thought 
facilitates recognition of items in memory that are correlated, 
\textit{i.e.} that share properties, which in turn provides more 
ingredients for analytic thought. Many species can learn, imitate, 
remember, and perhaps even form concepts. So this capacity to focus 
or defocus may be the key step in the attainment of complex language, 
religion, science, art, and other aspects of culture that make us 
unique. Defocused thought forges connections amongst items that share 
a deep structure but are superficially unrelated. Since these 
relations are often difficult to find, cultural learning also plays a 
key role here, filling in the missing links that the child does not 
find on its own (perhaps the vast majority). Because most of the 
fruits of such creative achievements are culturally transmitted, it 
is not necessary that we each individually generate an interconnected 
worldview from scratch. A hard-to-come-by concept or idea need only 
be realized in the mind of one individual; the other members of a 
society get it `for free' without any particularly focused or 
defocused thought.\\
\indent  As episodes accumulate in memory, and 
concepts acquire more states, the set of concepts grows and becomes 
more connected. When the same set of concepts arise repeatedly, 
concepts themselves form more abstract concepts, giving rise to 
semantic hierarchies. The probability increases that any experience 
triggers an episode encoded in memory which in turn triggers another, 
and so forth, generating a stream of thought. Streams of thought 
facilitate the formation of yet more abstract concepts.
 When the 
ratio of cognitive states to concepts reaches approximately 2:1, the 
probability becomes almost inevitable that concepts `join' in the 
sense that there come to exist cognitive states that are states of 
more than one concept. At this point, for any episode in the dominant 
subset of episodes there exists a possible route, or sequence of 
associative pathways (however indirect) to any other there exists a 
possible associative path (however indirect) to any other, enabling 
new experiences to be framed in terms of previous ones. We became 
able to encode, interpret, and reinterpret episodes using 
constellations of concepts of varying degrees of abstraction, wherein 
the meaning of each concept in the constellation shifts depending on 
the context provided by other simultaneously evoked concepts. 
Together they form a conceptual web that from the inside is 
experienced as an integrated model of the world, or worldview. It has 
been suggested that it is the conceptually integrated worldview that 
constitutes the replicating unit of cultural evolution. Like a living 
organism, a worldview may naturally be drawn to gaps or 
inconsistencies  and attempt to heal them (Gabora 1999, 2008). 
Elaboration of this approach may provide formal models of such 
psychological phenomena as repression, fragmentation, and 
integration.\\
\\\

\noindent{\textbf{\large{Acknowledgments}\\}}
This research is supported by grants from the Social Sciences and Humanities Research Council of 
Canada and the Research Council of the Free University of Brussels, Belgium. Part of the paper was written during a year spent by the first author at the PACE Center, Tufts University.\\

\section*{Appendix 1}
A closure relation is extensive, monotone and idempotent. This means that for $X, Y \subset 
{\cal X}$ we have
\begin{eqnarray} \label{eq:extensive}
&X \subset 
cl(X) \quad {\rm extensive} \\ \label{eq:monotone}
&X \subset Y 
\Rightarrow cl(X) \subset cl(Y) \quad {\rm monotone} \\ 
\label{eq:idempotent}
&cl(cl(X))=cl(X) \quad {\rm 
idempotent}
\end{eqnarray}
Proof: Consider an episode $x \in X 
\subset {\cal X}$. Since $x \leftrightarrow x$ this episode is 
connected to an episode $x \in {\cal X}$, hence, $x \in cl(X)$. This 
proves (\ref{eq:extensive}).\\
\indent Let us prove 
(\ref{eq:monotone}). Suppose there exist two subsets of episodes $X, 
Y \subset {\cal X}$ such that $X \subset Y$, and an episode $x \in 
cl(X)$. Since $x \in cl(X)$ this means that there exists an episode 
$y \in X$ such that $x \leftrightarrow y$. From $X \subset Y$ it 
follows that $y \in Y$. Hence, there exists an episode $y \in Y$ such 
that $x \leftrightarrow y$, which proves that $x \in cl(Y)$. Hence, 
this shows that $cl(X) \subset cl(Y)$. This proves 
(\ref{eq:monotone}).\\
\indent Let us prove (\ref{eq:idempotent}). 
 From (\ref{eq:extensive}) we know that $X \subset cl(X)$, and 
applying (\ref{eq:monotone}) we get $cl(X) \subset cl(cl(X))$. This 
means that to prove (\ref{eq:idempotent}) we only need to show that 
$cl(cl(X)) \subset cl(X)$. Consider $x \in cl(cl(X))$. This means 
that there exists an episode $y \in cl(X)$ such that $x 
\leftrightarrow y$. Since $y \in cl(X)$, there exists an episode $z 
\in X$ such that $y \leftrightarrow z$. Applying the transitivity of 
the `connection relation' we get that $x \leftrightarrow z$, which 
means that $x \in cl(X)$. Hence, we have shown that $cl(cl(X)) \subset 
cl(X)$, and we have proven (\ref{eq:idempotent}).\\.

\section*{Appendix 2}
\indent Let us prove that the closure 
relation conserves full connectedness.
%% start change (take away a point)t L
%% stop change L
We start by stating that a subset of episodes 
$X \subset {\cal X}$ is fully connected if 
\begin{equation}
\forall\ 
x, y \in X\ {\rm we\ have}\ x \leftrightarrow 
y
\end{equation}

\bigskip
\noindent
{\bf Proposition:} {\it 
$cl(\{x\})$ is a fully connected set of 
episodes}.

\bigskip
\noindent
Proof: Consider $y, z \in cl(\{x\})$, 
then we have $y \leftrightarrow x$ and $z \leftrightarrow x$, and 
hence from the transitivity property of the `connection relation' 
follows that $y \leftrightarrow z$.

\bigskip
\noindent
{\bf 
Proposition:} {\it If $X$ is a fully connected set of episodes, then 
also $cl(X)$ is a fully connected set of 
episodes}.

\bigskip
\noindent
Proof: Consider $x, y \in cl(X)$, then 
there exists $z, t \in X$ such that $x \leftrightarrow z$ and $y 
\leftrightarrow t$. Since $X$ is a fully connected set, we have $z 
\leftrightarrow t$. Hence, from the transitivity of the `connection 
relation' it follows that $x \leftrightarrow t$, and therefore $x 
\leftrightarrow y$. This proves that $cl(X)$ is fully 
connected.

\bigskip
\noindent
Hence, we have proven that the closure 
relation conserves full connectedness. \\

\section*{Appendix 3}
{\bf Proposition:} {\it If $X$ is a 
fully connected set of episodes, then $cl(X)$ is the biggest fully 
connected subset containing $X$}.

\bigskip
\noindent
Proof: Consider 
an arbitrary fully connected subset $Y$ such that $X \subset Y$. Let 
us prove that $Y \subset cl(X)$. Consider $x \in Y$ and an arbitrary 
$y \in X$. Since $Y$ is fully connected we have that $x 
\leftrightarrow y$. This means that $x \in cl(X)$. Hence, we have 
proven that $Y \subset cl(X)$.

\section*{Appendix 4}
{\bf Proposition:} {\it Consider $x, 
y \in {\cal X}$ two episodes. For $cl(\{x\})$ and $cl(\{y\})$ we 
have}
\begin{equation}
cl(\{x\})=cl(\{y\}) \quad {\rm or} \quad 
cl(\{x\}) \cap cl(\{y\})=\emptyset
\end{equation}
Proof: Suppose that 
$cl(\{x\}) \cap cl(\{y\})\not=\emptyset$, which means that there 
exists an episode $z \in cl(\{x\})$ and $z \in cl(\{y\})$. Hence, we 
have $z \leftrightarrow x$ and $z \leftrightarrow y$. From this it 
follows that $x \leftrightarrow y$. But then $x \in cl(\{y\})$, and 
hence, $\{x\} \subset cl(\{y\})$. Applying (\ref{eq:monotone}) we get 
$cl(\{x\}) \subset cl(\{y\})$. Analogously it can be shown that 
$cl(\{y\}) \subset cl(\{x\})$, and as a consequence 
$cl(\{x\})=cl(\{y\})$. This proves the proposition, namely that for 
the closure of two single episodes, the closure subsets are equal, or 
their intersection is empty.\\

\section*{Appendix 5}
We now prove 
that ${\cal X}$ is equal to the disjoint union of fully connected 
subsets, such that each of the subsets is the closure of a single 
episode.

\bigskip
\noindent
{\bf Proposition:} {We 
have
\begin{equation}
{\cal X}=\bigcup_{i \in I}X_i \quad X_i\ {\rm 
fully\ connected} \quad X_j \cap X_k=\emptyset\ {\rm for}\ j \not= 
k
\end{equation}
such that for each $X_i$ there exists a single 
episode $x_i \in {\cal X}$, such that $X_i = 
cl(\{x_i\})$}.

\bigskip
\noindent
Proof: From set theory it follows 
that ${\cal X}=\cup_{x \in {\cal X}}\{x\}$, and since for each $x \in 
{\cal X}$ it is the case that $\{x\} \subset cl(\{x\})$ then also 
${\cal X}=\cup_{x \in {\cal X}}cl(\{x\})$. Consider the set the set 
$\{cl(\{x\})\ \vert x \in {\cal X}\}$. From the above proposition it 
follows that if $cl(\{y\})$ and $cl(\{z\})$ are two elements of this 
set, $cl(\{y\})=cl(\{z\})$ or $cl(\{y\}) \cap cl(\{z\})=\emptyset$. 
This proves the proposition.\\

\section*{References}
\begin{description}
\item Aerts, D. 
(1982). Description of many physical entities Without the paradoxes 
encountered in quantum mechanics. \textit{Foundations of Physics, 
12,} 1131-1170. 
\item Aerts, D. (1983). Classical theories and non 
classical theories as a special case of a more general theory, 
\textit{Journal of Mathematical Physics, 24,} 2441-2453.

\item Aerts, D. (2007). General quantum modeling of combining concepts: A quantum field model in Fock space. Archive address and link: http://arxiv.org/abs/0705.1740.

\item Aerts, D. (2009). Quantum structure in cognition. Journal of Mathematical Psychology (submitted).

Aerts, D., Broekaert, J. and Smets, S. (1999a). The liar paradox in a quantum mechanical perspective. {\it Foundations of Science, 4}, pp. 115-132.

Aerts, D., Broekaert, J., Smets, S. (1999b). A quantum structure description of the liar paradox. {\it International Journal of Theoretical Physics, 38}, pp. 3231-3239.

\item Aerts, 
D. \& Gabora, L. (2005a). A theory of concepts and their combinations 
I: The structure of the sets of contexts and properties. 
\textit{Kybernetes, 34}(1\&2), 167-191.
\item Aerts, D. \& Gabora, L. 
(2005b). A theory of concepts and their combinations II: A Hilbert 
space representation. \textit{Kybernetes, 34}(1\&2), 192-221.
\item 
Aiello, L. \& Dunbar, R. (1993). Neocortex size, group size, and the 
evolution of language. \textit{Current Anthropology, 34}, 
184-193.
\item Ashby, F. G., \& Ell, S. W. (2002). Single versus 
multiple systems of learning and memory. In J. Wixted \& H. Pashler 
(Eds.) \textit{Stevens? handbook of experimental psychology: Volume 4 
Methodology in experimental psychology}. New York: Wiley.
\item 
Barkow, J., Cosmides, L., \& Tooby, J. (1992). \textit{The adapted 
mind.} New York: Oxford University Press.
\item Bar-Yosef, O. (1994). 
The contribution of southwest Asia to the study of the origin of 
modern humans. In M. Nitecki \& D. Nitecki (Eds.) \textit{Origins of 
anatomically modern humans}, Plenum Press.
Bouissac, P. (1998). 
Converging parallels: Semiotics and psychology in evolutionary 
perspective. \textit{Theory and Psychology}, 8(6), 731-753.

\item 
Beltrametti, E. \& Cassinelli, G. (1981). \textit{The logic of 
quantum mechanics.} Boston: Addison-Wesley.
\item Bickerton, D. (1990). \textit{Language and species.} Chicago: University of Chicago 
Press.

\item 
Barsalou, L.W. (1982). Context-independent and context-dependent information in concepts. \textit{Memory \& Cognition, 10}, 82-93.

\item Bruza, P. \& Cole, R. (2005). Quantum logic of semantic 
space: An exploratory investigation of context effects in practical 
reasoning. In S. Artemov, H. Barringer, A. S. d'Avila Garcez, L. C. 
Lamb and J. Woods (Eds.) \textit{We Will Show Them: Essays in Honour 
of Dov Gabbay}. College Publications, 1, pp. 339?361.
Bruza, P. 
Kitto, K. Nelson, D. \& McEvoy, C. (in press). Is there something 
quantum-like about the human mental lexicon? Journal of Mathematical 
Psychology. 
\item Busemeyer, J. R., Wang, Z., \& Townsend, J. T. 
(2006). Quantum dynamics of human decision making. \textit{Journal of 
Mathematical Psychology, 50}, 220-242.

\item Busemeyer, J. R. , Matthew, M., \& Wang, Z. (2006)  A Quantum Information Processing Theory Explanation of Disjunction Effects. {\it Proceedings of the Cognitive Science Society}.

\item Busemeyer, J. R. \& 
Wang, Z. (2007). Quantum Information Processing Explanation for 
Interactions between Inferences and Decisions. \textit{Proceedings of 
the AAAI Spring Symposium on Quantum Interaction} (pp. 91-97). 
Stanford University, March 26-28.

\item Carey, S. (1985). 
\textit{Conceptual change in childhood}. Cambridge: MIT Press.
\item 
Churchland, P. S. \& Sejnowski, T. (1992). \textit{The computational 
brain.} Cambridge MA: MIT Press.
\item Corballis, M. C. (1991). 
\textit{The lopsided ape: Evolution of the generative mind.} 
Cambridge:  Cambridge University Press.
\item Davis, J., \& 
Rovee-Collier, C. (1983). Alleviated forgetting of a learned 
contingency in 8-week-old infants. \textit{Developmental Psychology, 
19}, 353-365.
\item Dawkins, R. (1976). \textit{The selfish gene}. 
Oxford: Oxford University Press.
\item Deacon, T. (1997). \textit{The 
symbolic species}. New York: Norton. [p. 751] 
\item Dennett, D. 
(1978). \textit{Brainstorms}. Cambridge, MA: MIT Press.
\item 
Dikranjan, D., Giuli, E. \& Tozzi, A. (1988). Topological categories 
and closure operators. \textit{Questiones Mathematics, 11}(3), 
323-337. 
\item Donald, M. (1991) \textit{Origins of the modern 
mind.} Cambridge MA: Harvard University Press.
\item Edelman, G. E. 
(1993). \textit{Bright Air, Brilliant Fire: On the Matter of Mind.} 
New York: Basic Books.
\item Erdos, P. \& Renyi, A. (1959). 
\textit{On the random graphs}. 1(6), Institute of Mathematics, 
University of Debrecenians, Debrecar, Hungary.
\item Erdos, P. \& 
Renyi, A. (1960). \textit{On the evolution of random graphs}. 
Institute of Mathematics, Hungarian Academy of Sciences Publication 
number 5.
\item Fauconnier, G. \& Turner, M. (2002). \textit{The way 
we think: Conceptual blending and the mind's hidden complexities.} 
New York: Basic Books.
\item Gabora, L. (1998). A tentative scenario 
for the origin of culture. \textit{Psycoloquy, 9}(67).
\item Gabora, 
L. (1999). Weaving, bending, patching, mending the fabric of reality: 
A cognitive science perspective on worldview inconsistency. 
\textit{Foundations of Science, 3}(2), 395-428. 
\item Gabora, L. 
(2000). Conceptual closure: Weaving memories into an interconnected 
worldview. In G. Van de Vijver and J. Chandler (Eds.) 
\textit{Closure: Emergent Organizations and their Dynamics. Annals of 
the New York Academy of Sciences 901}, 42-53.
\item Gabora, L. 
(2002a). The beer can theory of creativity. In P. Bentley \& D. Corne 
(Eds.) \textit{Creative Evolutionary Systems,} (pp. 147-161). San 
Francisco CA: Morgan Kauffman.
\item Gabora, L. (2002b). Cognitive 
mechanisms underlying the creative process. In T. Hewett and T. 
Kavanagh (Eds.) \textit{Proceedings of the Fourth International 
Conference on Creativity and Cognition,} (pp. 126-133). October 
13-16, Loughborough University, UK. 
\item Gabora, L. (2003). 
Contextual focus: A cognitive explanation for the cultural transition 
of the Middle/Upper Paleolithic. In (R. Alterman \& D. Hirsch, Eds.) 
\textit{Proceedings of the 25th Annual Meeting of the Cognitive 
Science Society}, Boston MA, 31 July-2 August. Lawrence Erlbaum 
Associates.
\item Gabora, L. (2004). Ideas are not replicators but 
interrelated networks of them are. \textit{Biology and Philosophy, 
19}(1), 127-143.
\item Gabora, L. (2005). Creative thought as a 
non-Darwinian evolutionary process. \textit{Journal of Creative 
Behavior, 39}(4), 65-87.
\item Gabora, L. (2006). Self-other 
organization: Why early life did not evolve through natural 
selection. \textit{Journal of Theoretical Biology, 241} (3), 
443-450.
\item Gabora, L. (2008). The cultural evolution of socially 
situated cognition. \textit{Cognitive Systems Research, 9}(1-2), 
104-113.
\item Gabora, L. \& Aerts, D. (2002). Contextualizing 
concepts using a mathematical generalization of the quantum 
formalism. \textit{Journal of Experimental and Theoretical Artificial 
Intelligence, 14}(4), 327-358.
\item Gabora, L. \& Aerts, D. (2005). 
Evolution as context-driven actualization of potential: Toward an 
interdisciplinary theory of change of state. 
\textit{Interdisciplinary Science Reviews, 30}(1), 69-88.
\item 
Gabora, L., Rosch, E., \& Aerts, D. (2008). Toward an ecological 
theory of concepts. \textit{Ecological Psychology, 20}(1), 
84-116.
\item G\"ardenfors, P. (2000). \textit{Conceptual Spaces: The 
geometry of thought.} Cambridge, MA: MIT press. 
\item Gelman, R. 
(1993). A rational-constructivist account of early learning about 
numbers and objects. In D. Medin (Ed.) \textit{Learning and 
motivation,} (pp. 61-96). New York: Academic Press.
\item Hampton, S. 
J. (2004). Domina mismatches, scruffy engineering, exaptations and 
spandrels. \textit{Theory and Psychology 14}(2), 147-166.
\item 
Hancock, P. J. B., Smith, L. S. \& Phillips, W. A. (1991). A 
biologically supported error-correcting learning rule. {\it Neural 
Computation, 3}(2), 201-212.
\item Holden, S.B. \& Niranjan, M. 
(1997). Average-case learning curves for radial basis 
function
networks. {\it Neural Computation, 9}(2), 441-460.
\item 
Johnson-Laird, P. N. (1983). {\it Mental models.} Cambridge MA: 
Harvard University Press.
\item Kanerva, P. (1988). \textit{Sparse, 
distributed memory.} Cambridge MA: MIT Press.
\item Karmiloff-Smith, 
A. (1990). Constraints on representational change: Evidence from 
children?s drawing. \textit{Cognition, 34}, 57-83.
\item 
Karmiloff-Smith, A. (1992). \textit{Beyond modularity: A 
developmental perspective on cognitive science}. Cambridge MA: MIT 
Press.
\item Kauffman, S. (1993). \textit{Origins of order}. Oxford: 
Oxford University Press.
\item Keil, F. C. (1995). The growth of 
causal understandings of natural kinds. In S. Sperber, D. Premack, \& 
A. Premack (Eds.) \textit{Causal cognition: A multidisciplinary 
debate,} (pp. 234-262). Cambridge: Oxford University Press.
\item 
Klein, R. (1989). Biological and behavioral perspectives on modern 
human origins in South Africa. In P. Mellars \& C. Stringe (Eds.) 
\textit{The human revolution}. Edinburgh: Edinburgh University Press. 

\item Langton, C. G. (1992) Life at the edge of chaos. In (C. G. 
Langton, C. Taylor, J. D. Farmer \& S. Rasmussen, eds.) 
\textit{Artificial life II}. Boston: Addison-Wesley.
\item Leakey, R. 
(1984). \textit{The origins of humankind}. New York: Science Masters 
Basic Books.
\item Leslie, A. (2000). \textit{Theory of mind as a 
mechanism of selective attention}. Rutgers University Tech Report 
061.
\item Lieberman, P. (1991). \textit{Uniquely human: The 
evolution of speech, thought, and selfless behavior.} Boston: Harvard 
University Press.
\item Lu, Y.W., Sundararajan, N. \& Saratchandran, 
P. (1997). A sequential learning scheme for function approximation 
using minimal radial basis function neural networks. {\it Neural 
Computation, 9}(2), 461-478.
\item Maturana, H. \& Varela, F. 
(1973/1980). \textit{Autopoeisis and cognition: The realization of 
the living}. Boston: Reidel.
\item Matzel, L., Collin, C., \& Alkon, 
D. (1992). Biophysical and behavioral correlates of memory storage: 
Degradation and reactivation. \textit{Behavioral Neuroscience, 106}, 
954-963.
\item McBrearty, S. \& Brooks, A. S. (2000). The revolution 
that wasn?t: A new interpretation of the origin of modern human 
behavior. \textit{Journal of Human Evolution 39}(5), 453-563.
\item 
Mednick, S. (1962). The associative basis of the creative process. 
\textit{Psychological Review, 69,} 220?232.
\item Mellars, P. (1973). 
The character of the middle-upper transition in south-west France. In 
C. Renfrew (Ed.) \textit{The explanation of culture change}. London: 
Duckworth. 
\item Mellars, P. (1989a). Technological changes in the 
Middle-Upper Paleolithic transition: Economic, social, and cognitive 
perspectives. In P. Mellars \& C. Stringer (Eds.) \textit{The human 
revolution}.  Edinburgh: Edinburgh University Press. 
\item Mellars, 
P. (1989b). Major issues in the emergence of modern humans. 
\textit{Current Anthropology, 30}, 349-385.  
\item Mikkulainen 
(1997). Neural network perspectives on cognition and adaptive 
robotics. In (A Brown, Ed.) \textit{Natural language processing with 
subsymbolic neural networks.} Bristol UK, Philadelphia: Institute of 
Physics Press.
\item Mithen, S. (1996). \textit{The prehistory of the 
mind: A search for the origins of art, science, and religion.} 
London: Thames \& Hudson. 
\item Mithen, S. (1998). A creative 
explosion? Theory of mind, language, and the disembodied mind of the 
Upper Paleolithic. In  S. Mithen (Ed.) \textit{Creativity in human 
evolution and prehistory}. New York: Routledge.
\item Morowitz, H. J. 
(2002). \textit{The Emergence of Everything: How the World Became 
Complex.} Oxford University Press, New York.  
\item Neisser, U. 
(1963). The multiplicity of thought. \textit{British Journal of 
Psychology, 54}, 1-14.
\item Nelson, D. L. \& McEvoy, C. L. (2007). 
Activation at a distance is not spooky anymore: The conjunction of 
entangled associative structures and context. \textit{Proceedings of 
the AAAI Spring Symposium on Quantum Interaction,} Stanford 
University, March 26-28.
\item Nelson, D. L., McEvoy, C. L. \& 
Pointer, L. (2003). Spreading activation or spooky action at a 
distance. \textit{Journal of Experimental Psychology: Learning, 
Memory, and Cognition 29,} 42?51.
\item Osherson , D. N. \& Smith, E. 
E. (1981). On the adequacy of prototype theory as a theory of 
concepts. \textit{Cognition, 9}, 35-58.
\item Piaget 1926). 
\textit{The language and thought of the child}. Hillsdale, NJ: 
Lawrence Erlbaum.
\item Piron, C. (1976). \textit{Foundations of 
Quantum Physics.} Reading, MA: W.A. Benjamin.
\item Rosch (1978). 
Principles of categorization. In Rosch, E. \& Lloyd, B. (Eds.) 
\textit{Cognition and categorization}.  Hillsdale, NJ: Lawrence 
Erlbaum.
\item Rovee-Collier, C., Hartshorn, K., \& DiRubbo, M. 
(1999). Long-term maintenance of infant memory. \textit{Developmental 
Psychobiology, 35}, 91-102.
\item Ruff, C., Trinkaus, E. \& Holliday, 
T. (1997). Body mass and encephalization in Pleistocene Homo. 
\textit{Nature, 387,} 173-176.
\item Sloman, S. (1996). The empirical 
case for two systems of Reasoning. \textit{Psychological Bulletin, 
9}(1), 3-22.
\item Smolensky, P. (1988). On the proper treatment of 
connectionism. \textit{Behavioral and Brain Sciences, 11,} 
1?43.
\item Soffer, O. (1994). Ancestral lifeways in Eurasia---The 
Middle and Upper Paleolithic records. In M. Nitecki \& D. Nitecki, 
(Eds.) \textit{Origins of anatomically modern humans.} New York: 
Plenum Press. 
\item Spelke, E. S. (2000). Core knowledge. 
\textit{American Psychologist, 55}(11), 1233-1243.
\item Spoor, F. 
Leakey, M. G., Gathogo, P. N., Brown,  F. H., Anton, S. C., 
McDougall, I., Kiarie, C., Manthi, F. K. and Leakey, L. N. (2007). 
Implications of new early homo fossils from ileret, east of lake 
turkana, kenya. \textit{Nature, 448,} 688-691.
\item Stringer, C. \& 
Gamble, C. (1993). \textit{In search of the Neanderthals.} London: 
Thames \& Hudson. 
\item Thompson, E.(2007). \textit{Mind in Life - 
Biology, Phenomenology and the Sciences of Mind.} Cambridge MA: 
Harvard University Press.
\item Vetsigian, K., Woese, C., \& 
Goldenfeld, N. (2006). Collective evolution and the genetic code. 
\textit{Proceedings of the New York Academy of Science USA, 103,} 
10696?10701. 
\item W?chtersh?user, G. (1992). Groundwork for an 
evolutionary biochemistry: the iron-sulfur world. \textit{Prog. 
Biophys. Molec. Biol. 58,} 85-201.
\item Weber, B. H. (1998). 
Emergence of life and biological selection from the perspective of 
complex systems dynamics. In: G. van de Vijver, S. N. Salthe \& M. 
Delpos (Eds.) \textit{Evolutionary Systems: Biological and 
Epistemological Perspectives on Selection and Self-Organization.} 
Kluwer, Dordrecht. 
\item Weber, B. H. (2000). Closure in the 
emergence and evolution of life: Multiple discourses or one? In: J. 
L. R. Chandler \& G. Van de Vijver (Eds.) \textit{Closure: Emergent 
Organizations and their Dynamics, Annals of the New York Academy of 
Sciences 901,} pp. 132-138.
\item Weber, B.H., \& Depew, J.D. (1996). 
Natural selection and self-organization. \textit{Biology and Philosophy, 11}(1), 33-65.
\item White, R. (1982). Rethinking the 
Middle/Upper Paleolithic transition. \textit{Current Anthropology, 
23}, 169-189.
\item White, R. (1993). Technological and social 
dimensions of 'Aurignacian-age' body ornaments across Europe. In H. 
Knecht, A. Pike-Tay, \& R. White (Eds.) \textit{Before Lascaux: The 
complex record of the Early Upper Paleolithc.} New York: CRC 
Press.
\item Widdows, D. S. (2003). Orthogonal negation in vector 
spaces for modeling word meanings and document retrieval. In 
\textit{Proceedings of the 41st Annual Meeting of the Association for 
Computational Linguistics} (p. 136-143), Sapporo Japan, July 
7-12.
\item Widdows, D. \& Peters, S. (2003). Word vectors and 
quantum logic: experiments with negation and disjunction. In 
\textit{Eighth Meeting of the Association for the Mathematics of 
Language} (p. 141-154), Bloomington IN: Association for Computational 
Linguistics.
\item Williams, R. J. P. \& Frausto da Silva, J. J. R. 
(1999). \textit{Bringing Chemistry to Life: From Matter to Man.} 
Oxford University Press, Oxford.
\item Williams, R. J. P. \& Frausto 
da Silva, J. J. R. (2002). The systems approach to evolution. 
\textit{Biochem. Biophys. Res. Comm. 297,} 689-699. 
\item Williams, 
R. J. P. \&Frausto da Silva, J. J. R. (2003). Evolution was 
chemically constrained. \textit{J. Theor. Biol. 220,} 323-343.
\item 
Willshaw, D. J. \& Dayan, P. (1990). Optimal plasticity from matrix 
memory: What goes up must come down. {\it Journal of Neural 
Computation, 2}, 85-93.
\item Woese, C.R. (2002). On the evolution of 
cells. \textit{Proceedings of the National Academy of 
Sciences, 
99}(13), 8742-8747. 
\end{description}
\end{document}